\documentclass[achemso,aip,jcp,preprint,superscriptaddress,floatfix]{revtex4-1}
\usepackage[utf8]{inputenc}
\usepackage{color}
\usepackage{graphicx}
\usepackage{mathtools}
\usepackage{multirow}
\usepackage{amssymb}
\usepackage{palatino} 

\usepackage[english]{babel}
\usepackage{amsmath}
\usepackage{amsfonts}
\usepackage{graphicx}

\usepackage{algorithm}
\usepackage{algpseudocode}

\usepackage{xargs} 
\usepackage{lipsum}  
\usepackage[colorinlistoftodos,prependcaption,textsize=tiny]{todonotes}
\newcommandx{\addcite}[2][1=]{\todo[linecolor=red,backgroundcolor=red!25,bordercolor=red,#1]{#2}} 
\newcommandx{\checkthis}[2][1=]{\todo[linecolor=blue,backgroundcolor=blue!25,bordercolor=blue,#1]{#2}} 
\newcommandx{\addsome}[2][1=]{\todo[linecolor=green,backgroundcolor=green!25,bordercolor=green,#1]{#2}} 
\newcommandx{\modifythis}[2][1=]{\todo[linecolor=cyan,backgroundcolor=cyan!25,bordercolor=cyan,#1]{#2}}

\usepackage{threeparttable}
\usepackage{tikz}
\usetikzlibrary{shapes.geometric, arrows}
\tikzstyle{tensor}=[circle,draw=blue!50,fill=blue!20,thick]

\begin{document}
\title[]{Portably parallel construction of a CI wave function from a matrix-product state using the Charm++ framework}

\author{Wang Ting}
\affiliation{
Computer Network Information Center, Chinese Academy of Sciences, Beijing 100190, China
}
\affiliation{
Center of Scientific Computing Applications \& Research, Chinese Academy of Sciences, Beijing 100190, China
}
\affiliation{
School of Computer Science, Shaanxi Normal University, Xi'an 710119, China 
}


\author{Yingjin Ma}
\email{yingjin.ma@sccas.cn}
\affiliation{
    Computer Network Information Center, Chinese Academy of Sciences, Beijing 100190, China
}
\affiliation{
    Center of Scientific Computing Applications \& Research, Chinese Academy of Sciences, Beijing 100190, China
}

\author{Lian Zhao}
\affiliation{
    Computer Network Information Center, Chinese Academy of Sciences, Beijing 100190, China
}
\affiliation{
    Center of Scientific Computing Applications \& Research, Chinese Academy of Sciences, Beijing 100190, China
}

\author{Jinrong Jiang}
\affiliation{
    Computer Network Information Center, Chinese Academy of Sciences, Beijing 100190, China
}
\affiliation{
    Center of Scientific Computing Applications \& Research, Chinese Academy of Sciences, Beijing 100190, China
}

\date{\today}

\begin{abstract}        

The constructions of configuration interaction (CI) expansions from a matrix-product state (MPS) involves numerous matrix operations and the skillful sampling of important configurations when in a huge Hilbert space. 
In this work, we present an efficient procedure for constructing CI expansions from MPS using the {\sc Charm++} parallel programming framework,
upon which automatic load balancing and object migration facilities can be employed.
This procedure was employed in the MPS-to-CI utility (Moritz \textit{et al.}, \textit{J. Chem. Phys.} 2007, 126, 224109), sampling-reconstructed complete active space algorithm (SR-CAS, Boguslawski \textit{et al.}, \textit{J. Chem. Phys.} 2011, 134, 224101) and entanglement-driven genetic algorithm (EDGA, Luo \textit{et al.}, \textit{J. Chem. Theory Comput.} 2017, 13, 4699-4710).
It enhances productivity and allows the sampling programs evolve to their population-expansion versions (e.g., EDGA with population expansion [PE-EDGA]).
Examples of 1,2-dioxetanone and firefly dioxetanone anion (FDO$^-$) molecules demonstrated that 1) the procedure could be flexibly employed among various multi-core architectures;
2) the parallel efficiencies could be persistently improved simply by increasing the proportion of asynchronous executions;
3) PE-EDGA could construct a CAS-type CI wave function from a huge Hilbert space, with 0.9952 CI completeness and 96.7\% correlation energy via $\sim$1.66$\times$10$^6$ configurations (only 0.0000028\% of the total configurations) of a bi-radical state of FDO$^-$ molecule using the full valence active space within a few hours.

\end{abstract}

\maketitle

\section{Introduction}

To improve computational efficiency,
parallel application programming interfaces (APIs), such as the open multi-processing (OpenMP)\cite{LIAnalysis, OpenMP, JoseAnalysis, jin1999openmp} or message passing interface (MPI) \cite{LIAnalysis, Zheng2004FTC, articleJA, Negara2010Automatic, jin1999openmp}, 
are frequently employed in the field of computer science.
For quantum chemistry programs, the adoption of suitable APIs can accelerate code developments as well as maximize computing resources. 
The most common parallel schemes utilize OpenMP and/or MPI, such as the large-scale parallel configuration interaction (CI) by Knecht \textit{et al.}\cite{Knecht2008Large} and the hugely parallel multi-configurational self-consistent field (MCSCF) by Vogiatzis \textit{et al.} \cite{VogiatzisPushing} 
Recently, the compute unified device architecture \cite{Takizawa2010CheCUDA, articleJA, HawickParallel} and open computing language programming framework on heterogeneous devices have also been introduced in the quantum chemistry community. \cite{ufimtsev2008quantum, asadchev2012new, snyder2016gpu, kussmann2017employing} For example, the graphics processing unit (GPU)-accelerated complete active space self-consistent field (CASSCF) was recently employed in a study of excited-state molecular dynamics by Snyder \textit{et al.} \cite{snyder2016gpu} Until now, parallel implementations of various SCF routines (e.g., Hartree-Fock SCF, CASSCF, and DFT) turn to common functions in most of the popular quantum chemistry programs. Specific programs such as NWChem (designed for high-performance computing [HPC])\cite{HanwellOpen, ZBH2014NWChem, ValievNWChem} and PetaChem (designed for streaming processors such as GPU)\cite{IsbornExcited, UfimtsevGraphical, UfimtsevQuantum} has been developed to maximize the utilization of HPC clusters and GPU clusters, respectively. 

In strongly correlated systems, the parallel implementation of current deterministic approaches may be restricted because of the exponential expansions of configurations in the Hilbert space when extending the system size. Approximately 10$^{10}$ configurations on a single core should be the upper limit for these correlated calculations. \cite{holmes2016heat} For example, the well-known CASSCF approach, in which a superposition of all possible configurations within a given orbital subset (called the active space) is constructed, is limited to within 20 orbitals and electrons because of these exponential expansions.
In the last few decades, various approaches have been developed overcome the exponential expansions problem, including restricted or selected CI approaches,\cite{roos1990, ivanic2003direct, gidofalvi2008active, cleland2010communications, gagliardi2011, petruzielo2012semistochastic, li2013splitgas, evangelista2014adaptive, thomas2015stochastic, liu2016ici, coe2018machine, zimmerman2019evaluation} quantum Monte-Carlo approaches, \cite{tubman2016deterministic, tubman2018modern, Jin1999Mechanism} and reduced/renormalization approaches such as the density-matrix renormalization group (DMRG).\cite{schollwock2011, chan2011density, morokuma2013, ayers2012, chan2012, yana15, oliv15a, knec16a, brabec2020massively} 
At present, it is known that DMRG can handle active spaces with tens of electrons and orbitals while still maintaining a high accuracy close to full CI. \cite{LiuMultireference, Angelin}
Recently, selected CI approaches have contributed to renewed interest. A selected CI was shown to be able to attain a comparable accuracy with the DMRG for Cr$_2$ with relatively little computational cost. \cite{holmes2016heat, tubman2016deterministic, tubman2018modern} 
Comparing with the matrix-product states (MPS) wave function of DMRG, the deterministic wave function of the selected CI approach is typically considered simple and intuitive. As such, it can be employed as the reference wave function for multi-reference (MR) calculations. If all the important configurations are obtained in the reference wave function, then a quantitative MR result can be expected. \cite{luo2018externally, sharma2017semistochastic, tubman2018modern}

New hybrid approaches, such as the external-contracted MRCI (ec-MRCI) \cite{luo2018externally} and Epstein-Nesbet perturbation theory (ENPT) \cite{song2019multi}, have been developed successively to combine the advantages of the DMRG approach in acquiring static correlations with those of the selected CI approach in MR correlations. 
These approaches convert the wave function from its MPS form into its CI form (MPS-to-CI)
via the procedures discussed by Moritz \textit{et al.}, \cite{reiher2007} Boguslawski \textit{et al.}, \cite{boguslawski2011construction} and the authors of this paper.\cite{luo2017efficient}
The latter two sampling-evaluating-recording procedures for efficiently constructing a CAS-type wave function are practical. An early example is the Monte-Carlo-based sampling-reconstructed CAS (SR-CAS) algorithm, \cite{boguslawski2011construction} and another is the entanglement-driving genetic algorithm (EDGA).\cite{luo2017efficient} However, improving the computational efficiency of these procedures is highly desirable,  as a huge Hilbert space is normally employed in the samplings. 

Improving efficiency via paralleling is not typically a trivial task for modern computers, and several factors, such as portability, latency of communication, and load balance, must be carefully considered.
%
As numerous processors are asynchronously used for sampling, evaluating, and recording, load balance is a major issue when massively parallel systems are implemented. As such, the {\sc Charm++} is a potential API that delivers good performance as already shown in NAMD \cite{phillips2005scalable, nelson1996namd} and OpenAtom \cite{mandal2017parallel, jain2015charm++, jain2016openatom} programs.
%
{\sc Charm++} is a parallel programming system based on {\sc C++} and is built on a portable object programming model and supported by adaptive runtime system. \cite{Charm++Web, CHARM++:,
Kale2009Charm}  
It automatically performs dynamic load balancing of task parallel or data parallel applications through a separate load balancing strategy suite; this feature perfectly matches our asynchronously sampling-evaluating-recording process.
%

Using the {\sc Charm++} parallel programming system, we refactored the procedure of constructing CI expansions from an MPS wave function. This procedure can be employed as the kernel in other CAS-type wave function reconstructing strategies (Section II). The benchmark results of CI reconstruction of firefly lighting molecules were illustrated via various architectures, ranging from a laptop, workstations, to the two HPC clusters (Section III and Section IV). Our conclusions are presented in Section V.

\section{Implementing Details}

\subsection{MPS wave function and CI reconstruction}

In conventional CI theory, an arbitrary electronic state 
$|\psi \rangle$ spanned by the $L$ orbital is expressed as a linear combination of the occupation number vector (ONV) $|\sigma \rangle$, where the CI coefficients $c_{\sigma _{1}\cdots\sigma_{L}}$ are expressed as expansion coefficients:
\begin{equation}\label{eq_psi}
|\psi \rangle  =  \sum\limits_{\sigma}c_{\sigma}|\sigma \rangle = \sum \limits_{\sigma _{1}, \cdots, \sigma _{L}} c_{\sigma _{1}\cdots\sigma_{L}}| \sigma_{1}\cdots\sigma_{L}\rangle  
\end{equation}
The ground state $|\sigma\rangle$ has four possible occupied states for the first spatial orbit, which are $|\uparrow\downarrow\rangle$, $|\uparrow\rangle$, $|\downarrow\rangle$, and $|0\rangle$. When turning to MPS ansatz, the CI coefficients $c_{\sigma_{1}\cdots\sigma_{L}}$ can be coded as the product of the $m_{l-1}\times m_{l}$-dimensional matrix $M^{\sigma}$ = $\{M_{\alpha_{l-1}\alpha_l}^{\sigma_{l}}\}$:
\begin{equation}\label{eq_psiMPS}
|\psi \rangle = \sum \limits _{\sigma_{1}, \cdots, \sigma_{L}} \sum \limits _{\alpha_{1},\cdots,\alpha_{L-1}}  M_{1\alpha_{1}}^{\sigma_{1}} M_{\alpha_{1}\alpha_{2}}^{\sigma_{2}} \cdots M_{\alpha_{L-1}1}^{\sigma_{L}}|\sigma_{1}\cdots\sigma_{L} \rangle = \sum \limits_{\sigma} M^{\sigma_{1}} M^{\sigma_{2}} \cdots M^{\sigma_{L}}|\sigma\rangle  .
\end{equation}
Among them, the first and the last matrices are $1\times m_{l}$-dimension rows and $m_{L - 1}\times 1$-dimension column vectors, respectively. The summation on the $\alpha_{l}$ index is folded into a matrix-matrix multiplication to obtain the final equation.

Moritz \textit{et al.} \cite{reiher2007} proposed a method for determining all determinant weights in the DMRG scanning process in MPS ansatz. Here, all determinant weights are preserved in the first step, and the determinant weights of the iterative step change as the ground state changes in the actual calculation process. In such cases, the weight of a determinant state, which is also denoted as ONV, can be obtained as follow:
\begin{equation}\label{eq_mps2ci}
c_{\sigma_{1}\cdots\sigma_{L}} = M^{\sigma_{1}}[\sigma_{1}]M^{\sigma_{2}}[\sigma_{2}]\cdots M^{\sigma_{L}} [\sigma_{L}] .
\end{equation}
Among them, the $M$ matrix obtained by the base transformation is stored in the DMRG. We denote this as the MPS-to-CI procedure.

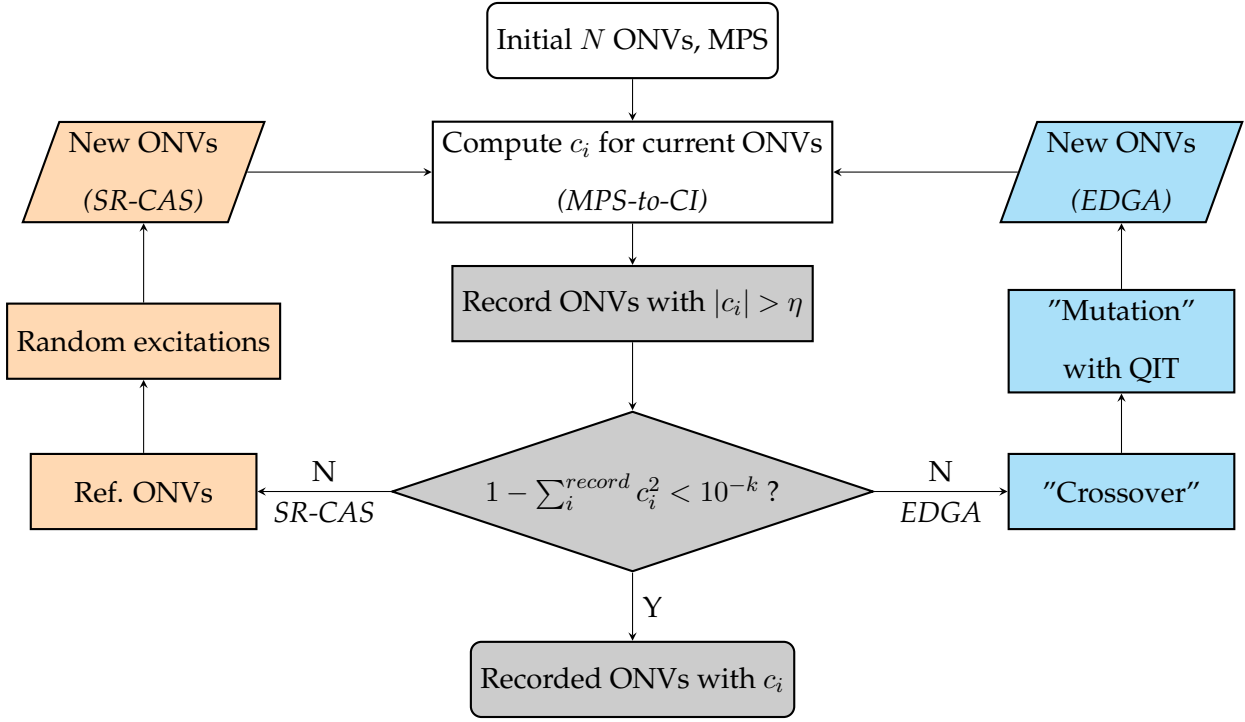
\begin{figure}[htb]	    
    \tikzstyle{startstop} = [rectangle, rounded corners, minimum width = 2cm, minimum height=1cm,text centered, draw = black,thick]
    \tikzstyle{startstop1} = [rectangle, rounded corners, minimum width = 2cm, minimum height=1cm,text centered, draw = black,thick,fill=black!20]
    \tikzstyle{io} = [trapezium, trapezium left angle=70, trapezium right angle=110, minimum width=2cm, minimum height=1cm, text centered,align=center, draw=black,thick,fill=cyan!30]
    \tikzstyle{io2} = [trapezium, trapezium left angle=70, trapezium right angle=110, minimum width=2cm, minimum height=1cm, text centered,align=center, draw=black,thick,fill=orange!30]
    \tikzstyle{process}  = [rectangle, minimum width=3cm, minimum height=1cm, text centered, draw=black, align=center,thick]
    \tikzstyle{process1}  = [rectangle, minimum width=3cm, minimum height=1cm, text centered, draw=black, align=center, draw = black,thick,fill=black!20]
    \tikzstyle{process2}  = [rectangle, minimum width=3cm, minimum height=1cm, text centered, draw=black, align=center, draw = black,thick,fill=orange!30]     
    \tikzstyle{process3}  = [rectangle, minimum width=3cm, minimum height=1cm, text centered, draw=black, align=center, draw = black,thick,fill=cyan!30]
    \tikzstyle{decision} = [diamond, aspect = 3, text centered, draw=black,thick,fill=black!20]
    \tikzstyle{arrow} = [->,>=stealth]
    \begin{tikzpicture}[node distance=1cm]
    \node[startstop](start){Initial $N$ ONVs, MPS};
    \node[process, below of = start, yshift = -0.75cm](in1){Compute $c_i$ for current ONVs \\ \textit{(MPS-to-CI)}};
    \node[process1, below of = in1 , yshift = -0.75cm](pro1){Record ONVs with $|c_i|$ $>$ $\eta$};
    \node[decision,below of = pro1, yshift = -1.5cm](dec1){$1-\sum_{i}^{record}c^{2}_{i}<10^{-k}$ ? };
    \node[process2,left of = dec1, xshift = -5.5cm, yshift = 0.0cm](pro21){Ref. ONVs};
    \node[process3,right of = dec1, xshift = 5.5cm, yshift = 0.0cm](pro22){"Crossover"};
    \node[process3,right of = pro1, xshift = 5.5cm, yshift = -0.50cm](pro32){"Mutation" \\ with QIT};
    \node[process2,left of = pro1, xshift = -5.5cm, yshift = -0.50cm](pro31){Random excitations};
    \node[io, right of = in1,  xshift = 5.5cm, yshift = 0cm](io1){New ONVs \\\textit{(EDGA)}};
    \node[io2, left of = in1,  xshift = -5.5cm, yshift = 0cm](io2){New ONVs \\\textit{(SR-CAS)}};
    \node[startstop1, below of = dec1, yshift = -1.5cm](out1){Recorded ONVs with $c_i$};
    \coordinate (point1) at (-3cm, -6cm);
    \draw [arrow] (start) -- (in1);
    \draw [arrow] (in1) -- (pro1);
    \draw [arrow] (pro1) -- (dec1);
    \draw [arrow] (dec1) -- node [right] {Y} (out1);
    \draw [arrow] (dec1) -- node [above] {N} node [below] {\textit{SR-CAS}}(pro21);
    \draw [arrow] (dec1) -- node [above] {N} node [below] {\textit{EDGA}} (pro22);
    \draw [arrow] (pro21) -- (pro31);
    \draw [arrow] (pro22) -- (pro32);
    \draw [arrow] (pro31) -- (io2);
    \draw [arrow] (pro32) -- (io1);
    \draw [arrow] (io1) -- (in1); 
    \draw [arrow] (io2) -- (in1); 
    \end{tikzpicture}	
    \caption{Flowchart of the MPS-to-CI, SR-CAS, and EDGA procedures.}\label{fig_three}
\end{figure}

Based on the MPS-to-CI procedure, SR-CAS and EDGA algorithms \cite{boguslawski2011construction, luo2017efficient} were developed, which aimed aimed to construct a reference deterministic wave function. With these algorithms, important configurations can be constructed with high priority therefore, the CASCI-type wave function could be obtained, and this wave function was used as the reference wave function in subsequent MR calculations. \cite{luo2018externally, song2019multi}
In the case of SR-CAS, a random sampling (Monte-Carlo) procedure was employed in three steps: 1) a random number between 1 and \textit{n} determined how many excitations were performed from a reference Slater determinant (SD, normally Hartree–Fock SD at the start); 2) two further random numbers within the interval of 1 and \textit{L} determined the spin-orbitals of the reference SD that was to be exchanged with one from the non-occupied orbitals; and 3) all excitation operators were applied to the reference SD, which then yielded a new trial SD in the representation of ONV.\cite{boguslawski2011construction} In the case of EDGA, the random sampling procedure was integrated as a ``mutation" process into a genetic algorithm, in which the CI expansions were obtained via both ``crossover" and ``mutation" processes.\cite{luo2017efficient} Here, the ``crossover" operation meant randomly selecting two SDs and generating a new SD as the combination of one’s alpha spin-orbitals and the other’s beta spin-orbitals with roulette selection correction. The ``mutation" operation was similar to that of SR-CAS, except that the probability of excitations was corrected via orbital entanglements obtained using the quantum information theory. \cite{legeza2004quantum, rissler2006measuring, boguslawski2012entanglement} These procedures are illustrated in Fig. \ref{fig_three}.

\subsection{Implementation using Charm++}

As shown in the previous subsection and the illustration in Fig. \ref{fig_three}, the MPS-to-CI procedure was the kernel component in the three CI reconstruction procedures. 
As such, we refactored the entire MPS-to-CI procedure using the {\sc Charm++} framework as the starting point.
The key elements in the MPS-to-CI refactoring procedure were 1) the chare array, 2) the pack/unpack (PUP) framework, and 3) optional reductions/callbacks.\cite{CHARM++:} 
A basic unit of parallel computation in the {\sc Charm++} programming framework is a chare,\cite{CHARM++:, Kale2009Charm} which is similar to a process or an actor. At its most basic level, a chare is a C++ object, and a single ONV object can be mapped to a chare. A {\sc Charm++} computation consists of a large number of chares distributed on a system’s available processors that interact with each other via asynchronous method invocations. 
Considering that there are a number of ONVs, the collection type of the (1-D) chare array was used to map the ONVs and benefit from {\sc Charm++} seed-based load balancing. 
Each chare in a chare array had a unique index of type \textit{ckarrayIndex} that was mapped to the local manager according to the corresponding ONV index. Array elements can be created and destroyed dynamically on any processing elements (PEs) and migrated between PEs, and the messages of elements can still arrive correctly. 
Since asynchronous invocations are used in the {\sc Charm++} program, the PUP framework is a general method for describing data in an object and uses a description for serialization. 
The use order of PUP is similar to {\sc C++}, which constructor objects, processes, and destroys them. 
In addition, {\sc Charm++} supports the reduction of array members, and callback functions are usually combined with reductions. For example, the build-in \textit{CkReduction::sum\_double} function is used when checking for CI completeness once all the CI coefficients of ONVs have been obtained. Reduction applies a single operation to data items scattered across multiple processors and collects the results in one location. However, the reductions/callbacks is not mandatory, because the satisfied ONVs and their CI coefficients can be recorded distributedly and simply via the same stream operations as those in {C++}. 

\begin{figure}[h]
    \label{figMPS2CI}    
    \tikzstyle{startstop} = [rectangle, rounded corners, minimum width = 2cm, minimum height=1cm,text centered, draw = black, fill = red!0]
    \tikzstyle{io} = [trapezium, trapezium left angle=80, trapezium right angle=100, minimum width=2cm, minimum height=1cm, text centered, draw=black, fill = blue!0]
    \tikzstyle{process} = [rectangle, minimum width=3cm, minimum height=1cm, text centered, draw=black, fill = yellow!0]
    \tikzstyle{decision} = [diamond, aspect = 3, text centered, draw=black, fill = green!0]
    \tikzstyle{arrow} = [->,>=stealth]
    \centering
    \begin{tikzpicture}[node distance=2.0cm]
        \node (start) [startstop] {MPS, aiming ONVs};
        \node (model) [below of=start,process] {Distributed among chare array in {\sc Charm++}};
        \node (chare0) [below of=model,process,xshift=-5.5cm] {PUP MPS};
        \node (chare1) [below of=model,process,xshift=-1cm] {PUP MPS};
        \node (chare2) [below of=model,process,xshift=3.5cm] {PUP MPS};
        \node (chare3) [below of=model,process,xshift=7cm ] {...};        
        \node (message0) [below of=chare0,io,xshift=0cm] {ONVs$^{1,...,i}$};
        \node (message1) [below of=chare1,io,xshift=0cm] {ONVs$^{j,...,k}$};
        \node (message2) [below of=chare2,io,xshift=0cm] {ONVs$^{l,...,m}$};
        \node (message3) [below of=chare3,io,xshift=0cm] {...};
        \node (cal0) [below of=message0,process,xshift=0cm] {Cal. each $c$};
        \node (cal1) [below of=message1,process,xshift=0cm] {Cal. each $c$};
        \node (cal2) [below of=message2,process,xshift=0cm] {Cal. each $c$};
        \node (cal3) [below of=message3,process,xshift=0cm] {...};                
        \node (stop) [below of=model,startstop,yshift=-6.5cm] {Collect $\sum c_i^2$ on all chares via the {\sc Charm++} reductions (Optional) };
        
        \draw [arrow] (start) -- (model);
        \draw [arrow] (model) -- (chare0);
        \draw [arrow] (model) -- (chare1);
        \draw [arrow] (model) -- (chare2);
        \draw [arrow] (model) -- (chare3);
        \draw [arrow] (chare0) -- (message0);
        \draw [arrow] (chare1) -- (message1);
        \draw [arrow] (chare2) -- (message2);
        \draw [arrow] (chare3) -- (message3);
        \draw [arrow] (message0) -- (cal0);
        \draw [arrow] (message1) -- (cal1);
        \draw [arrow] (message2) -- (cal2);
        \draw [arrow] (message3) -- (cal3);
        \draw [arrow] (cal0) -- (stop);
        \draw [arrow] (cal1) -- (stop);
        \draw [arrow] (cal2) -- (stop);
        \draw [arrow] (cal3) -- (stop);
    \end{tikzpicture}
    \caption{Workflow of the refactored MPS-to-CI procedure using the {\sc Charm++} framework} \label{mps2ci-charm}
\end{figure}
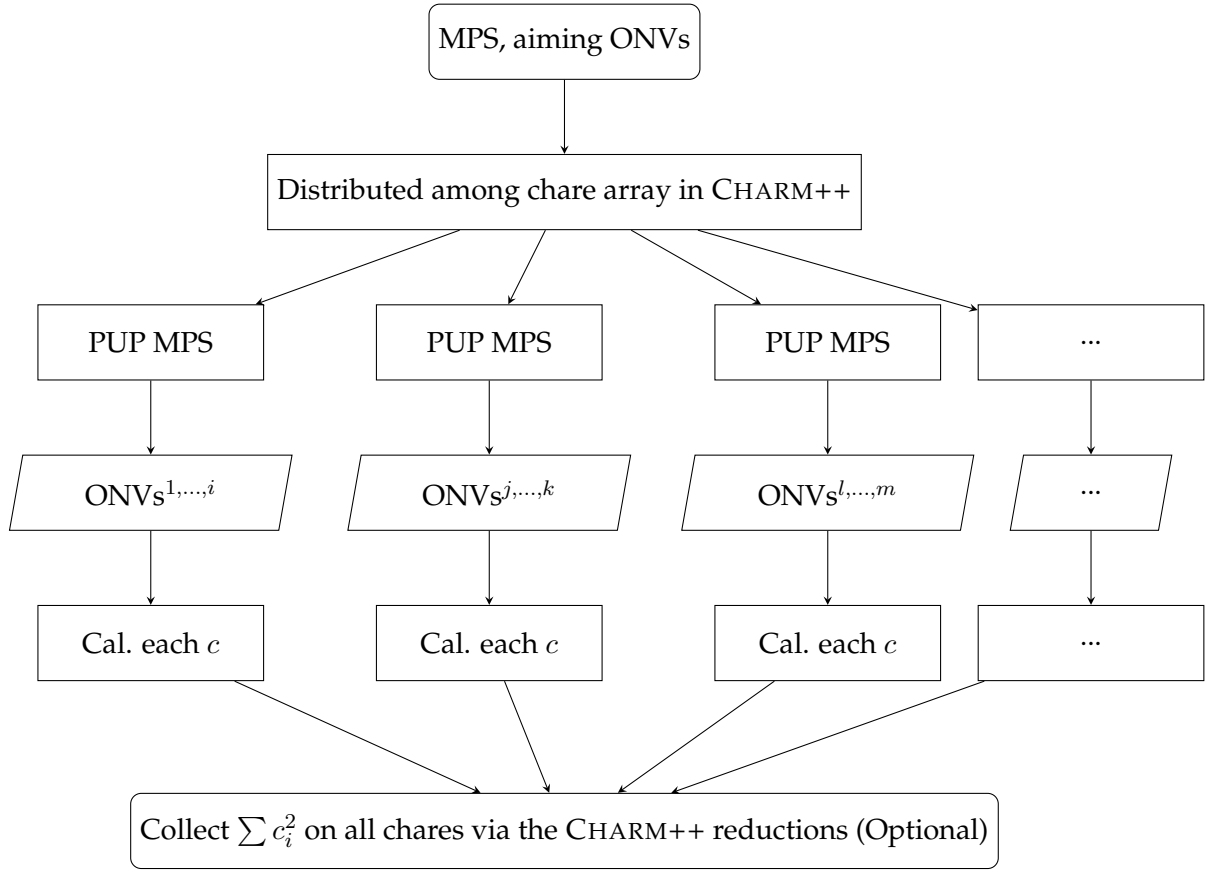

As the {\sc Charm++} uses a message-driven execution model, the MPS wave function should be passed as the message into each chare, so that it can be accessed by PEs. 
Here, the PUP framework is used to describe the MPS in an object and use that description for serialization. During implementing, all parameters in MPS should be packed and unpacked, including the $M$ matrix on each site, on a left, right, and local basis. 
After all related parameters in MPS have been passed as messages, the same algorithm illustrated by Moritz \textit{et al.} \cite{reiher2007} can be used in every chare with the corresponding ONVs, which can be accessed by \textit{ckarrayIndex}. Subsequently the calculated results can be collected using the mechanism of reductions, or recorded ``on-the-fly" for each \textit{ckarrayIndex}. The entire workflow for the refactored MPS-to-CI procedure is illustrated in Fig.\ref{mps2ci-charm}.

Once the MPS-to-CI procedure has been refactored, the SR-CAS and EDGA algorithms can be implemented using the \textit{do-while} or \textit{for-loop} structure in C++. All pseudo codes are listed in Algorithms \ref{pseudocode} and \ref{pseudocode3}.


\begin{algorithm}[H]
  \caption{The pseudo code of refactored MPS-to-CI, SR-CAS, and EDGA workflows.}
  \label{pseudocode}
   \begin{algorithmic}[1]
   \Require  MPS; ONVs; Nchares; Ent (optional);  
   \Ensure   ONVs with coefficients $c_{i}$    
   \State \#include \textit{Charm++ header}
   \State Load MPS, ONVs, Ent (optional)
   \State $MPS\_Charm = $ MPS  \Comment{PUP MPS, ONVs (\textit{italic} means it is employed in {\sc Charm++})} 
   \State $ONVs\_Charm = $ ONVs
   \State $DetCharm=CProxy\_DetCharmClass::ckNew(Nchares)$ \Comment{Allocate 1-D chare array} 
   \For{$iter = 0$ to $N$}
       \If {$iter == 0$} \Comment{MPS-to-CI procedure via chare array} 
         \State $DetCharm.mps2ci\_Charm(MPS\_Charm, ONVs\_Charm)$
       \Else
         \If {$SRCAS$} \Comment{SR-CAS procedure via chare array} 
           \State $DetCharm.srcas\_Charm(MPS\_Charm, ONVs\_Charm)$ 
         \Else  \Comment{EDGA procedure via chare array} 
           \State $DetCharm.edga\_Charm(MPS\_Charm, ONVs\_Charm,$ Ent$)$
         \EndIf 
       \EndIf
   \EndFor
   \end{algorithmic}
\end{algorithm}

\begin{algorithm}[H]
  \caption{The pseudo code for the MPS-to-CI, SR-CAS, and EDGA procedures. }
  \label{pseudocode3}
   \begin{algorithmic}[1] 
      \State void $mps2ci/srcas/edga\_Charm$($ONVs\_Charm$, $MPS\_Charm$, $Nchares$, Ent)
      \qquad \For{$i = 1$ to \textit{nONVs}/\textit{Nchares}} \Comment{Distribute ONVs}
       \State  int \textit{iposition}=\textit{i*Nchares};
       \State  $localONV=ONVs\_Charm[ckarrayIndex+ipos]$; \Comment{Specific ONV}
       \If {MPS-to-CI}
      \State  double $c_i$=$mps2ci\_kernel(localONV,MPS\_Charm)$ \Comment{Compute $c_i$ of current ONV}    
       \Else  
         \If {SRCAS} \Comment{If SR-CAS}
         \State  $localONV=localONV.random\_excitation$;   
         \Else  \Comment{If EDGA}
         \State  $localONV2=ONVs\_Charm[ckarrayIndex+iposition+shift]$; 
         \State  $localONV=localONV.alpha + localONV2.beta$;   \Comment{Crossover}
         \State  $localONV=localONV.mutation$(Ent); \Comment{Mutation}
         \EndIf      
       \State  double $c_i$=$mps2ci\_kernel(localONV,MPS\_Charm)$ \Comment{Compute $c_i$ of current ONV} 
       \EndIf  
      \qquad   \EndFor   
   \end{algorithmic}
\end{algorithm}

\subsection{EDGA with population expansion (PE-EDGA)}

By utilizing the automatic load balancing and object migration facilities in {\sc Charm++}, we improved the EDGA procedure to a population-expansion version (the so-called PE-EDGA). The parallel efficiencies can be persistently improved simply by increasing the proportion of asynchronous executions. The algorithm is presented as follows (the flowchart is illustrated in Fig. \ref{figPE-EDGA}):

1) Prepare the initial ONVs. Usually, using a set of ONVs calculated from SR-CAS is ideal because the random excitation in SR-CAS can ensure the genetic diversity of the ONVs.

2) Run the refactored EDGA procedure. Each ONV on the chare can be evolved basing on its own "crossover" and "mutation" operations via the chare-based random number generator, so that the loads on each chares are different. Note that {\sc Charm++} uses the asynchronous execution model for automatic load balancing and object migration facilities. Therefore, the satisfied ONVs and their CI coefficients ($c_i$) can be recorded on-the-fly, and the number of EDGA loops should be limited to conserve random access memory.  

3) Analyze the recorded ONVs. The recorded ONVs can be sorted, and the CI completeness can be calculated by $\sum c_i^2$. A similar hash data structure to one proposed by Boguslawski \textit{et al.}  \cite{boguslawski2011construction} can be employed for efficiency. All the recorded ONVs are used as the initial population of ONVs for the next cycle, so that it is population expansion.

4) If CI completeness is larger than a given threshold, stop the program; if not, proceed to step 2).

\begin{figure}[h] \label{figPE-EDGA}  
    \tikzstyle{startstop} = [rectangle, rounded corners, minimum width = 2cm, minimum height=1cm,text centered, draw = black,thick]
    \tikzstyle{io} = [trapezium, trapezium left angle=70, trapezium right angle=110, minimum width=2cm, minimum height=1cm, text centered,align=center, draw=black,thick,fill=cyan!0]
    \tikzstyle{process}  = [rectangle, minimum width=3cm, minimum height=1cm, text centered, draw=black, align=center,thick]
    \tikzstyle{decision} = [diamond, aspect = 3, text centered, draw=black,thick,fill=black!0]
    \tikzstyle{arrow} = [->,>=stealth]
    \centering
    \begin{tikzpicture}[node distance=2.0cm]
    \node[startstop](start){Initial population of ONVs};
    \node[process, below of = start, yshift = 0cm](in1){Several EDGA iterations in {\sc Charm++}};
    \node[process, below of = in1 , yshift = 0cm](pro1){Pupulation analysis in {\sc C++}};
    \node[decision,below of = pro1, yshift = -0.5cm](dec1){$1-\sum_{i}^{record}c^{2}_{i}<10^{-k}$ ? };
    \node[process,right of = dec1, xshift = 5.5cm, yshift = 0.0cm](pro22){Sorting and popution expansion};
    \node[io, right of = in1,  xshift = 5.5cm, yshift = 0cm](io1){New population};
    \node[startstop, below of = dec1, yshift = -0.5cm](out1){Recorded ONVs with $c_i$};
    \coordinate (point1) at (-3cm, -6cm);
    \draw [arrow] (start) -- (in1);
    \draw [arrow] (in1) -- (pro1);
    \draw [arrow] (pro1) -- (dec1);
    \draw [arrow] (dec1) -- node [right] {Y} (out1);align=center,
    \draw [arrow] (dec1) -- node [above] {N} (pro22);
    \draw [arrow] (pro22) -- (io1);
    \draw [arrow] (io1) -- (in1); 
    \end{tikzpicture}
    \caption{Workflow of PE-EDGA procedure.}   \label{figPE-EDGA}  
\end{figure}
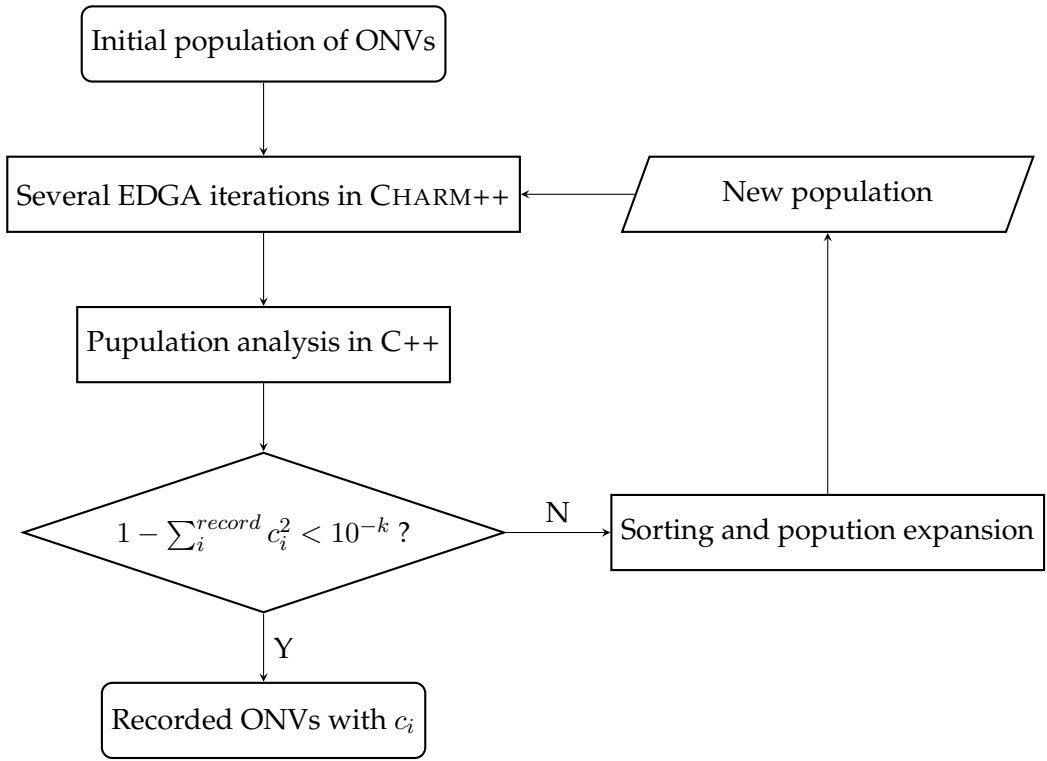

It should be noted that SR-CAS can also be adapted to its PE version simply by disabling the crossover operations in EDGA.  

\section{Computational details}

\begin{figure}[htbp]
	\includegraphics[scale=0.7, bb=0  0 400 250]{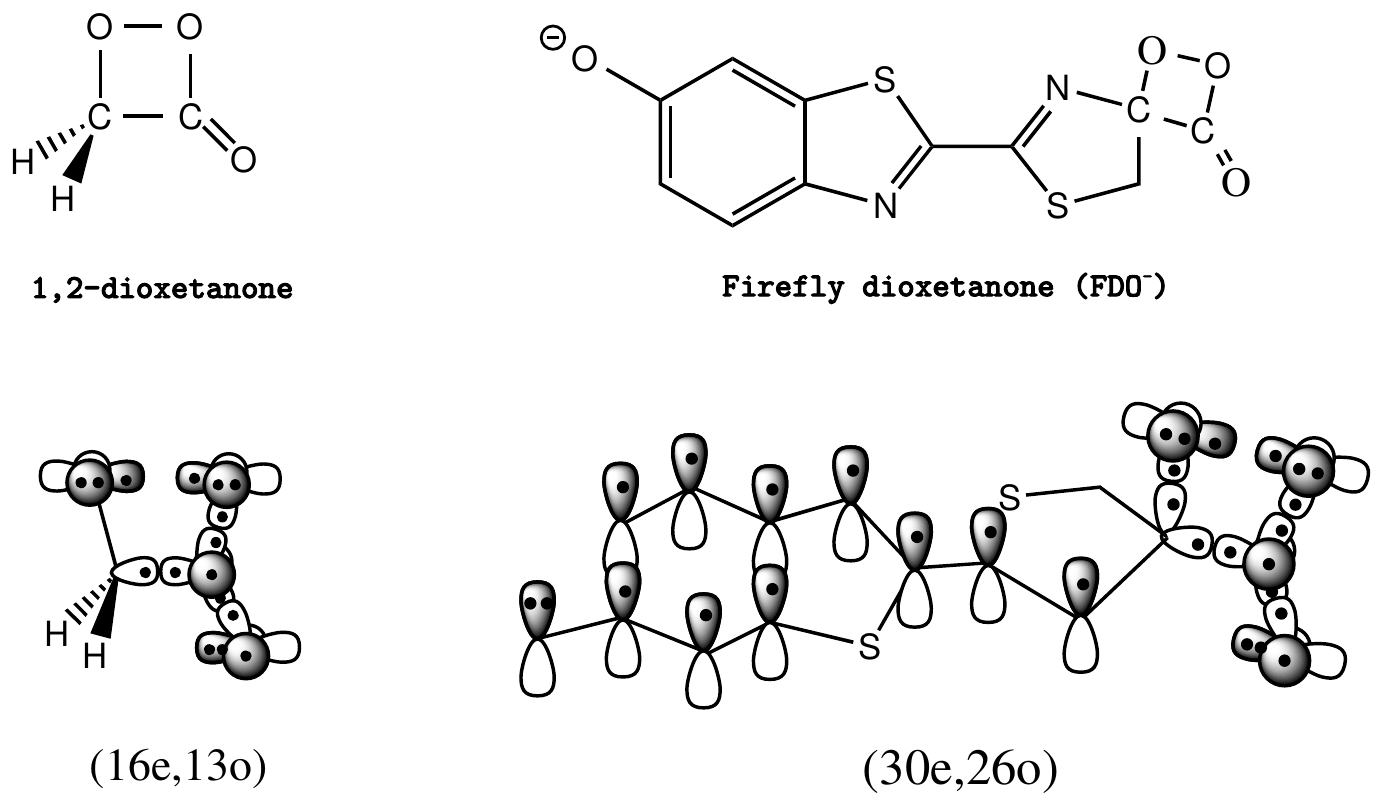}
	\caption{Illustration of 1,2-dioxetanone, FDO$^{-}$, and their active spaces.}
	\label{diofdo}
\end{figure}

The typical chemiluminescent molecules 1,2-dioxetanone, firefly dioxetanone anion (FDO$^{-}$) are used as benchmark systems as shown in Fig.\ref{diofdo} for the refactored CI wave function constructing procedures.
The state-averaged DMRG(16,13)[1000]-SCF and DMRG(30,26)[1000]-SCF calculations were implemented for lowest states of the S$_1$ minimum structure of 1,2-dioxetanone and the conical intersection structure of FDO$^{-}$, respectively. The structures of these molecules belong to the bi-radical region in the lighting process and were identified in our previous work.\cite{ma2020pccp}  All the relevant valence orbitals were involved in these calculations.
The ANO-RCC basis set with the VDZP contraction \cite{widm90, pier95} atomic orbital basis with scalar relativity and the second-order Douglas-Kroll-Hess Hamiltonian \cite{wolf02, reih04a, reih04b} were used. 
The DMRG package {\sc QCMaquis} \cite{kell15a, kell16} was used as the engine for the DMRG calculations, and the {\sc OpenMOLCAS}-{\sc QCMaquis} interface \cite{fdez2019openmolcas} performed the orbital optimization process. The refactored codes, which were made based on a developing version of {\sc QCMaquis}, were integrated into our self-written {\sc eMC} and {\sc NDQC} codes, updated in GitHub \cite{githubcode} and will be deployed at the biomed community \cite{biomed}.  
In the CAS ansatz, the number of possible ONVs was 1,656,369 and 59,693,548,345,600. The coefficients of all ONVs could be evaluated with MPS-to-CI for 1,2-dioxetanone, although this was not possible for FDO$^{-}$ using the full valence space (i.e., CAS[30,26]). 
In addition, the CAS(16,13) active space, which was simply truncated based on the natural orbital occupation number, was also used for FDO$^{-}$ for the evaluation.

Four typical multi-core architectures were used, including the following: 

1) Laptop; the DELL Inspiron-7591 with 6-core Intel i7-9750H CPU and 24GB MEM;

2) Workstation; the HUAWEI 2288H V5 with 24-core two-way Intel Xeon Sliver-4116 CPU and 128GB MEM;

3) HPC clusters-1; the Era cluster of Chinese Academy of Sciences with 24-core two-way Intel Xeon E5-2680V3 and 128GB MEM on each node;

4) HPC clusters-2; the HPC cluster of Chinese academy of sciences with 32-core Hygon 7185 processor and 128 GB MEM on each node.

When implementing the calculations using {\sc Charm++}, the number of cores (i.e., nPEs or nCores) and number of chares (i.e., nChares) should be specified at the same time.  

\section{Results}

\subsection{Parallel efficiency for the MPS-to-CI kernel}

\begin{figure}[htbp]
\begin{minipage}[t]{0.2\linewidth}
\centering
	\includegraphics[height=6.0cm, width=8.0cm]{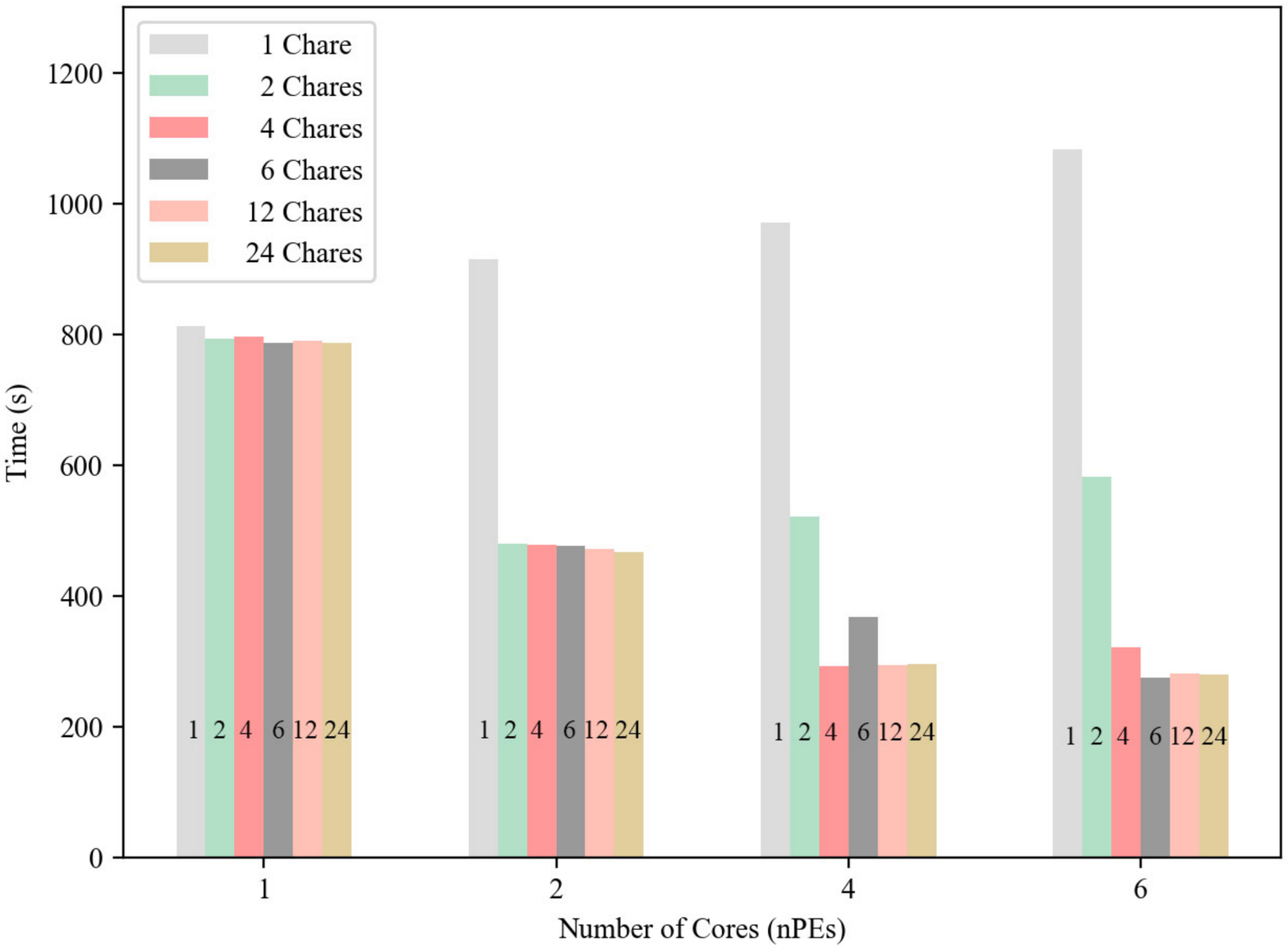}
	\end{minipage}
	\hfill
	\begin{minipage}[t]{0.5\linewidth}
	\centering
	\includegraphics[height=6.0cm,width=8.0cm]{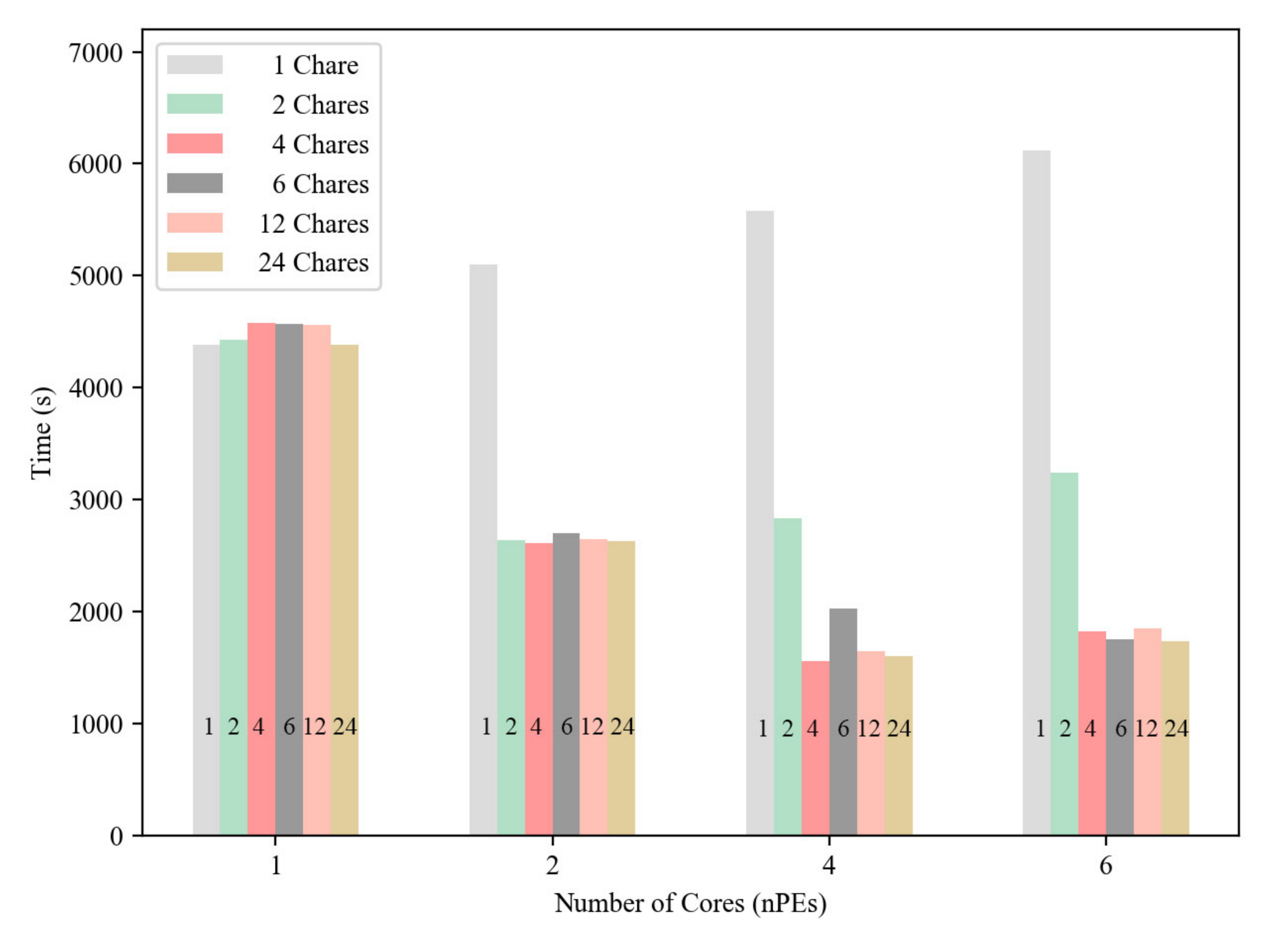}
	\end{minipage}
	\caption{\label{tab-pc} The number of cores (nPEs), nChares, and total wall time ($s$) in the refactored MPS-to-CI procedure for all ONVs in (16,13) active space using 1,2-dioxetanone (\textit{left}) and FDO$^{-}$ (\textit{right}) molecules in a laptop.} 
	\label{fig_four}
\end{figure}

\begin{figure}[htbp]
\begin{minipage}[t]{0.2\linewidth}
\centering
	\includegraphics[height=6.0cm, width=8.0cm]{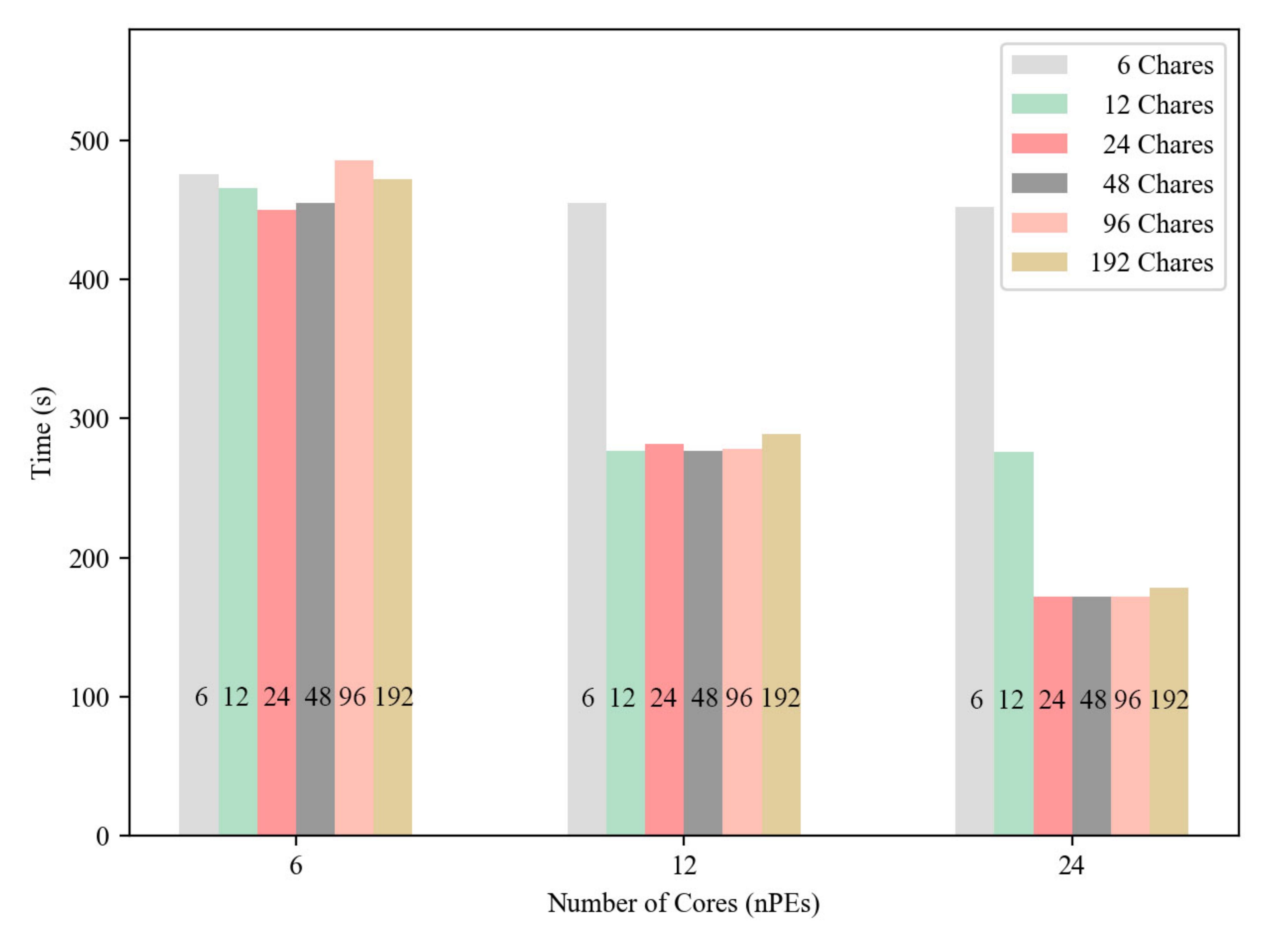}
	\end{minipage}
	\hfill
	\begin{minipage}[t]{0.5\linewidth}
	\centering
	\includegraphics[height=6.0cm,width=8.0cm]{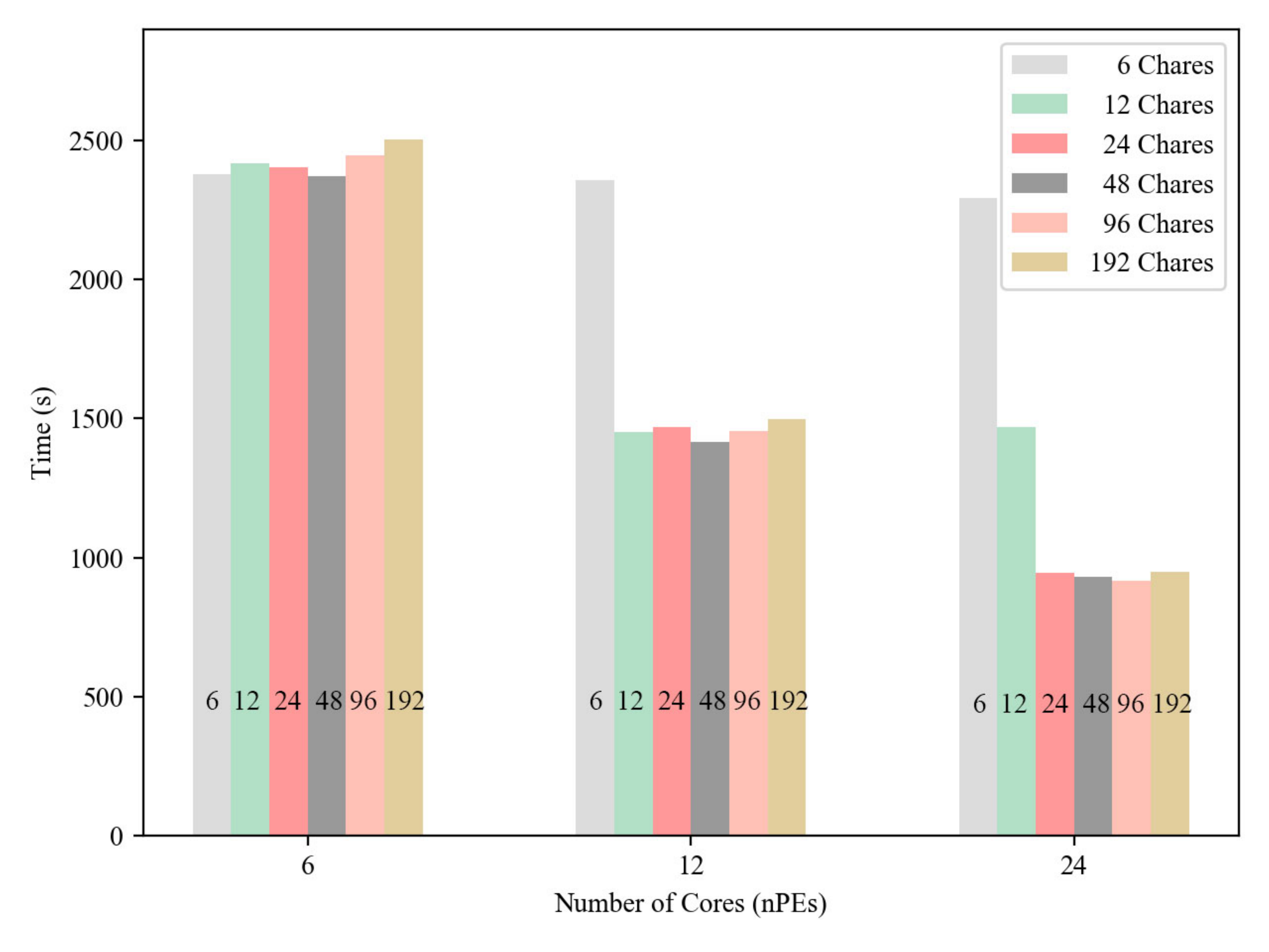}
	\end{minipage}
	\caption{\label{tab-ws} The number of cores (nPEs), nChares, and total wall time ($s$) for the refactored MPS-to-CI procedure for all ONVs in (16,13) active space using 1,2-dioxetanone (\textit{left}) and FDO$^{-}$ (\textit{right}) molecules in a workstation.} 
	\label{fig_five}
\end{figure}

\begin{figure}[htbp]
\begin{minipage}[t]{0.2\linewidth}
\centering
	\includegraphics[height=6.0cm, width=8.0cm]{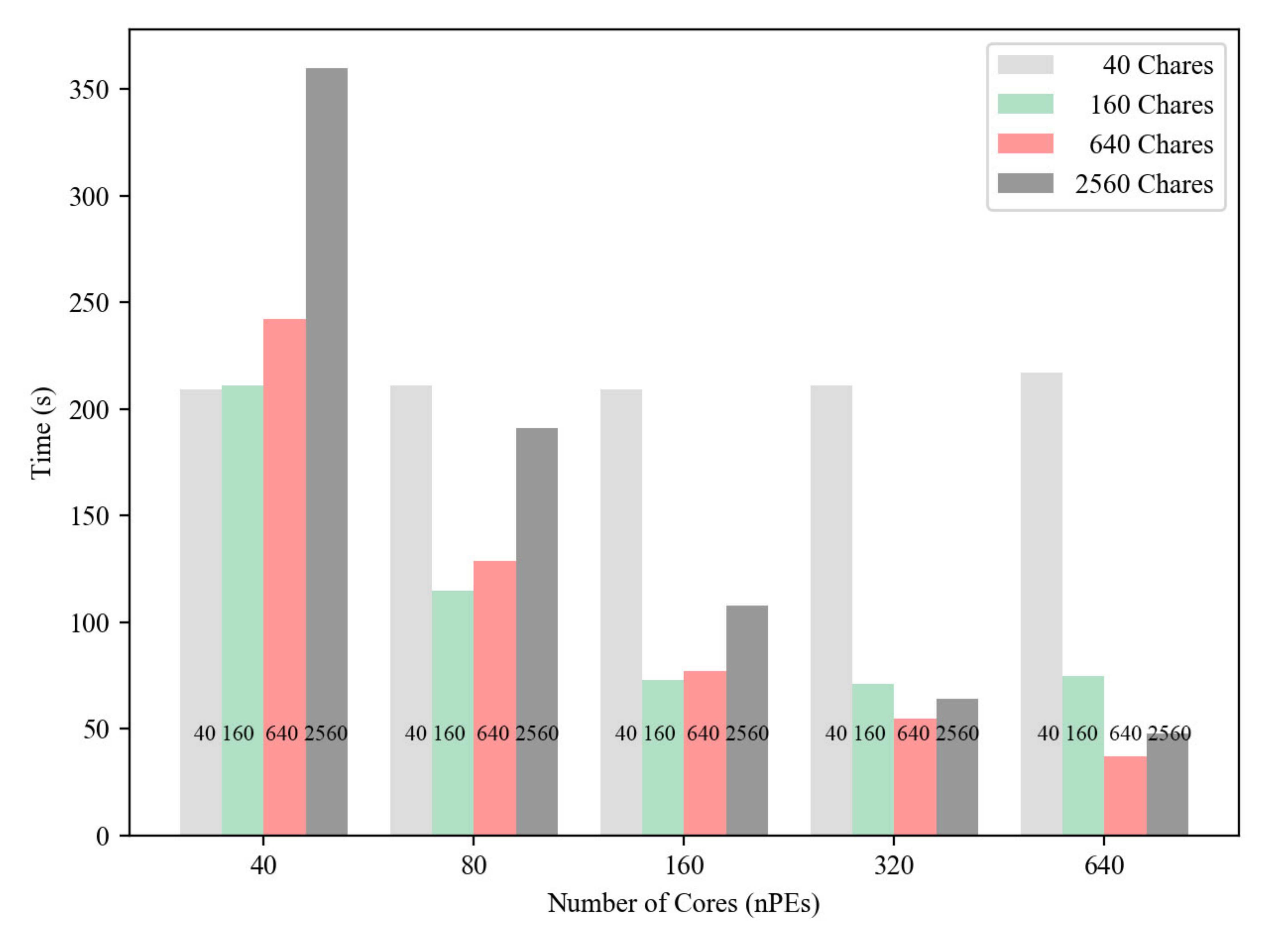}
	\end{minipage}
	\hfill
	\begin{minipage}[t]{0.5\linewidth}
	\centering
	\includegraphics[height=6.0cm,width=8.0cm]{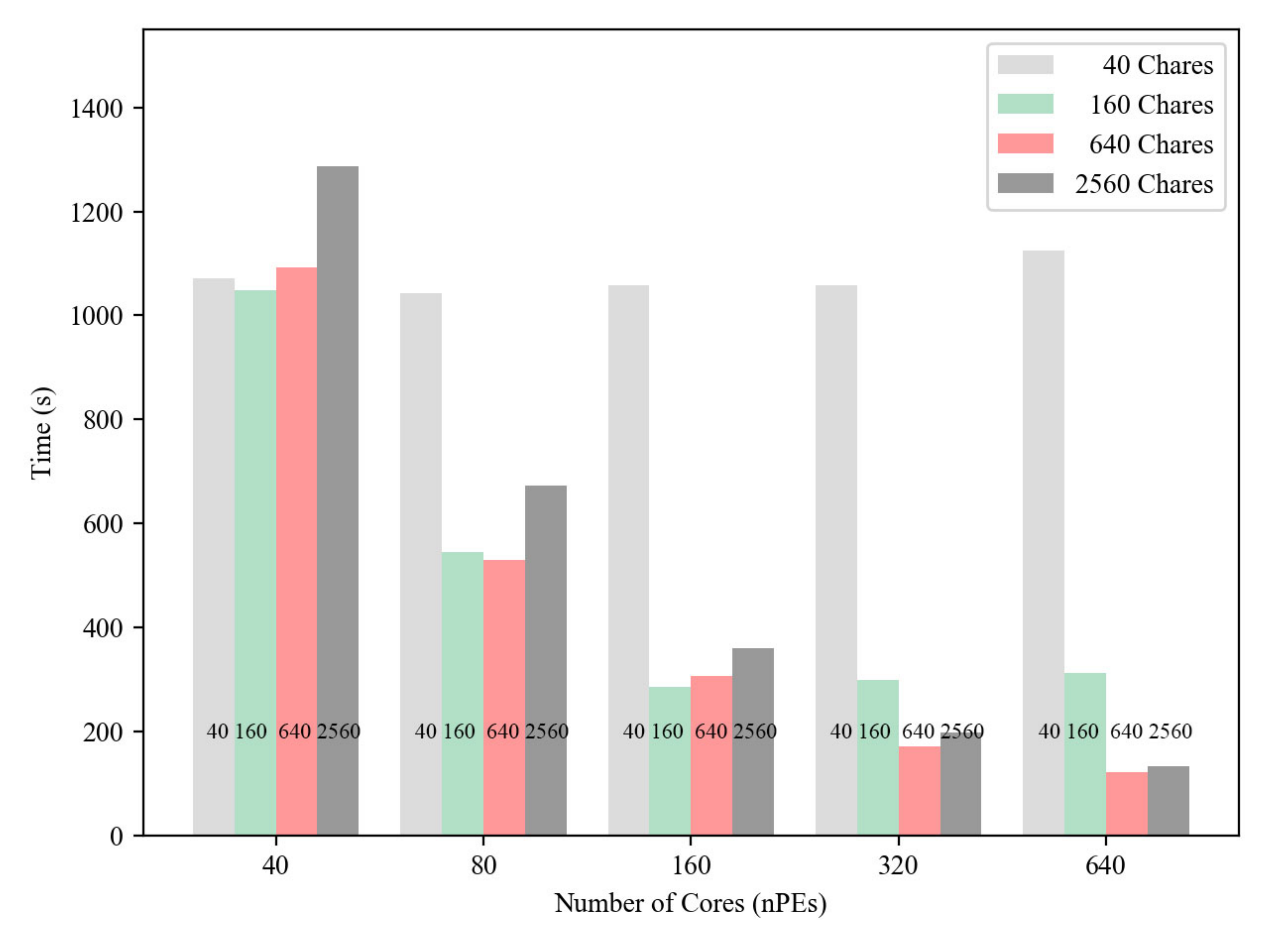}
	\end{minipage}
	\caption{\label{tab-hpc} The number of cores (nPEs), nChares, and total wall time ($s$) for the refactored MPS-to-CI procedure for all ONVs in (16,13) active space using 1,2-dioxetanone (\textit{left}) and FDO$^{-}$ (\textit{right}) molecules in HPC clusters-1.} 
	\label{fig_six}
\end{figure}

As illustrated in Fig. \ref{fig_three}, the MPS-to-CI procedure served as the kernel engine in constructing the deterministic CI wave functions; therefore, it was necessary to provide a detailed benchmark of its performance in various multi-core architectures. The ONVs in CAS(16,13) active spaces, which were the full valence for 1,2-dioxetanone and truncated space for FDO$^{-}$, were used as the benchmark efficiency in reconstructing CI wave functions.  
The results are listed in Fig. \ref{fig_four} (laptop), Fig. \ref{fig_five} (workstation), and Fig. \ref{fig_six} (HPC clusters-1).

It was discovered that the nChares had to be equal to or larger than the nPEs/nCores and with integer multiples to avoid hampering the efficiency. 
For the laptop (Fig. \ref{fig_four}), the parallel efficiencies were about 87\% (two cores), 70\% (four cores), 50\% (six cores) for both 1,2-dioxetanone and FDO$^{-}$.
The rapid decline in efficiency was mainly due to fluctuations in the CPU frequency (the max turbo frequency [$\thicksim$4.5 GHz] was employed for limited core[s] and the base frequency ($\thicksim$2.6 GHz) was employed for full cores.
The results of the workstation (Fig. \ref{fig_five}) were similar to the laptop; the parallel efficiencies for both were around 86\% (12 cores) and 68\% (24 cores) compared with the six-cores case.  
For the HP clusters (Fig. \ref{fig_six}), the parallel efficiencies were about 91\% (80 cores), 72\% (160 cores), 48\% (320 cores), and 36\% (640 cores) for 1,2-dioxetanone, and about 98\% (80 cores), 94\% (160 cores), 78\% (320 cores), and 56\% (640 cores) for FDO$^{-}$ when compared with the 40 cores. Fluctuations in the CPU frequency could be omitted for HPC clusters, and as such, a higher parallel efficiency was observed. In additionally, the parallel efficiency for FDO$^{-}$ was superior to that of 1,2-dioxetanone, which may be attributable to the data size when packing and unpacking. 
 
%

\begin{figure}[htbp]
	\centering
	\includegraphics[height=12.0cm,width=16.0cm]{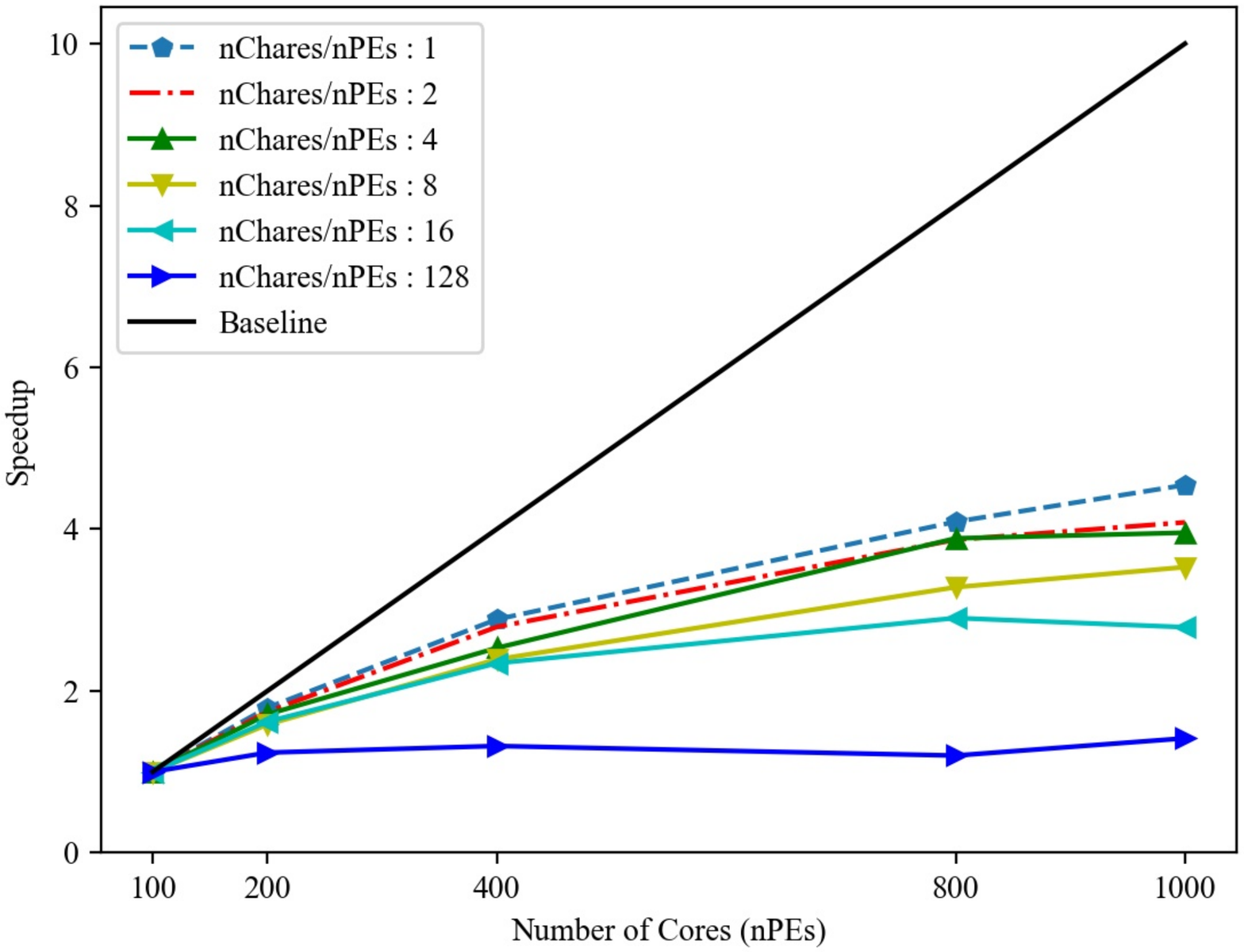}
	\caption{\label{tab-ws} Speedup for the implementation of the refactored MPS-to-CI procedure for all ONVs in (16,13) active space using FDO$^{-}$ molecules in HPC clusters.} 
	\label{fig_seven}
\end{figure}

In the results presented in Fig. \ref{fig_four}, \ref{fig_five}, \ref{fig_six}, it can be observed that the ratio between the nPEs/nCores and nChares greatly affected the efficiency (e.g., nChares should be equal or larger than the nPEs (ncores) and with integer multiples). To determine the optimum ratio between the nPEs (ncores) and nChares for the calculation, we picked several typical nChare:nPEs ratios (i.e. 1, 2, 4, 8, 16, and 128) and evaluated their performance ranging from 100 nPEs (ncores) to 1000 nPEs (ncores). The results are illustrated in Fig. \ref{fig_seven}. It was determined that the ratio of 1:1 showed the best parallel efficiency of about 43\% for 1000 nPEs (ncores) compared with that of 100 nPEs (ncores). In addition, the step-wise parallel efficiencies for 100--200, 200--400, and 400--800 were around 85\%, 75\%, and 65\%. This implied that the parallel efficiency was affected by the data distribution when a mess of nPEs (ncores) was employed. Here, it should be noted that the {\sc Charm++} dynamical load balance feature was not employed, as the imported ONVs were fixed. In the later PE-EDGA calculation, an obvious improvement in parallel efficiency was anticipated.   

\subsection{Constructing a CAS-type CI wave function for FDO$^-$}

After obtaining the optimized conditions for the MPS-to-CI engine, we began reconstructing a CI wave function for the FDO$^{-}$ molecule using the MPS from the DMRG(30,26)[1000]-SCF calculation.
The FDO$^{-}$ molecule has a conical intersection point; therefore, the non-dynamical electron correlation effect plays an important role in the description of near-degeneracy electronic states. The charge-transfer character, whereby the electron can be excited from far-end O$^-$ to the breaking C-C bond via the $\pi$ bridge, was also observed for the calculated state. 
In addition, there was no symmetry that could be used, and the number of ONVs could be up to around $6\times 10^{13}$. 
All these factors made CAS-type representations for the FDO$^{-}$ molecule highly challenging. 

\begin{figure}[htbp]
\begin{minipage}[t]{0.2\linewidth}
    \centering
	\includegraphics[height=7.0cm, width=8.0cm]{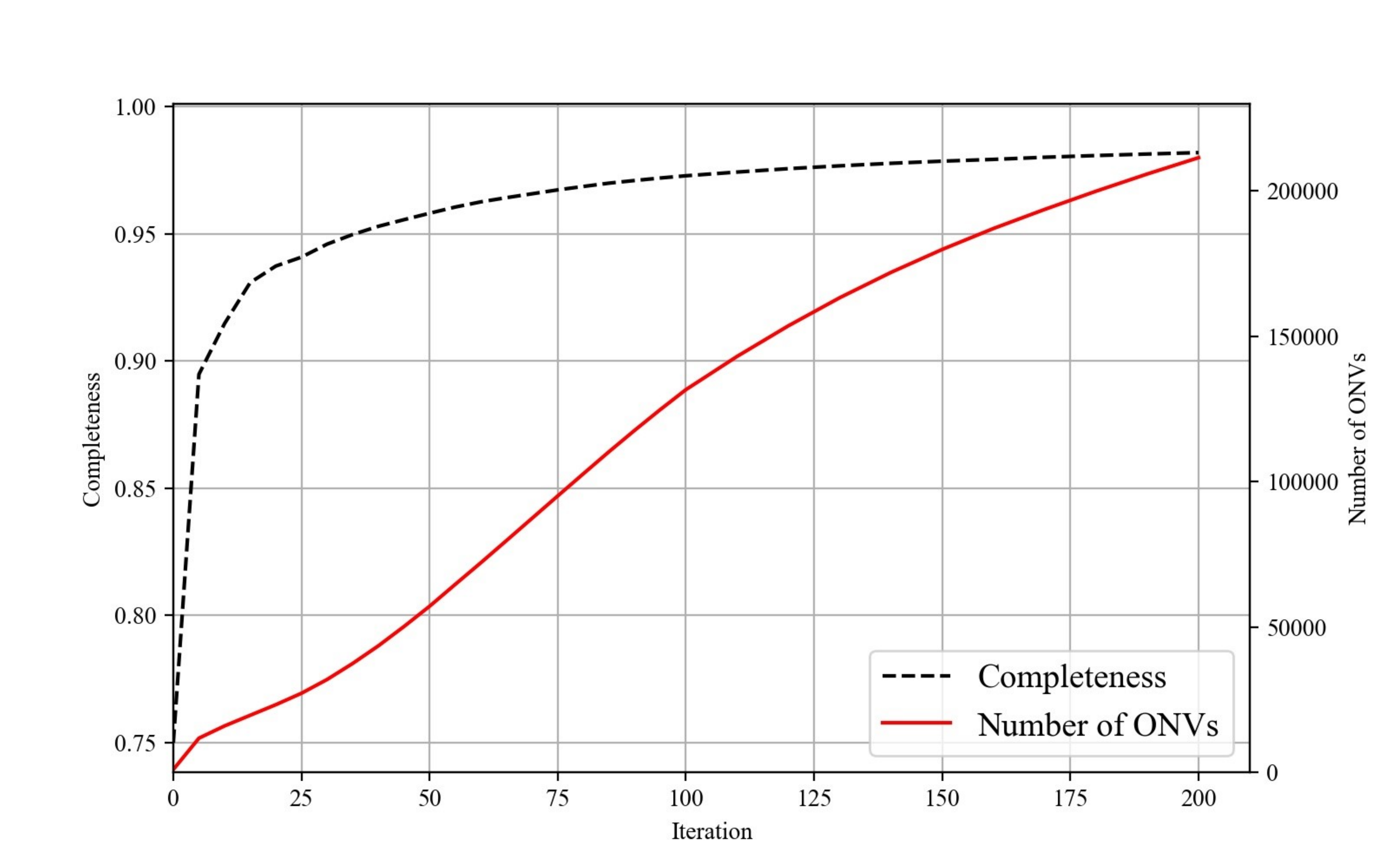}
	\end{minipage}
	\hfill
	\begin{minipage}[t]{0.5\linewidth}
	\centering
	\includegraphics[height=7.0cm,width=8.0cm]{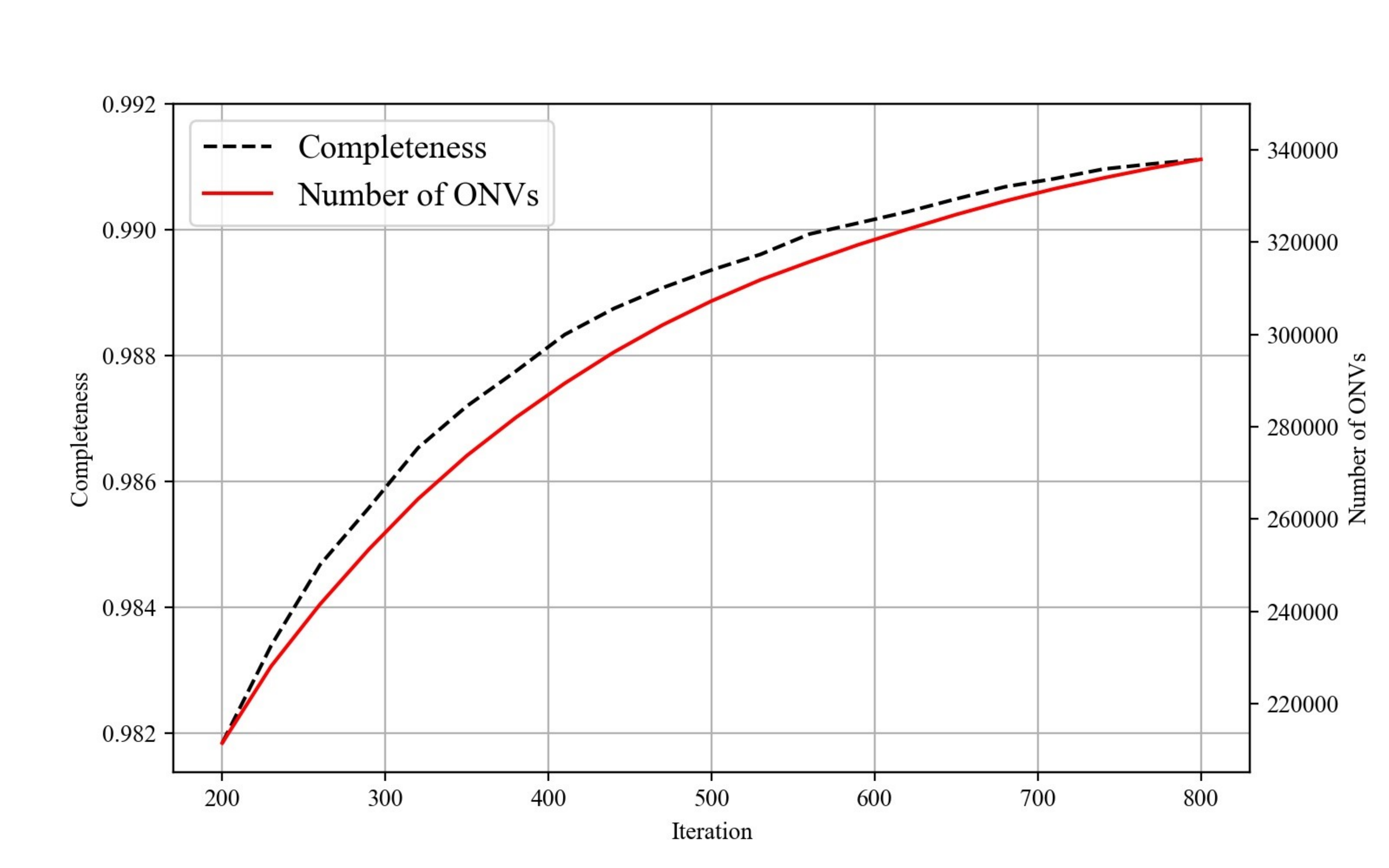}
	\end{minipage}
	\caption{The illustrations of CI completeness ($\sum c_i$), number of ONVs, and EDGA iterations during the PE-EDGA calculations with recording threshold $|c_i| >0.0001$.}\label{fig_eight}
\end{figure}

To obtain reliable CI representations for FDO$^{-}$, the PE-EDGA procedure (with 0.0001 of CI threshold for recording) was employed using laptop, workstation, and HPC clusters successively. 
The sampled ONVs from SR-CAS calculation were used as the initial ONVs, including 921 ONVs with a completeness of 0.75.
The first 20 PE-EDGA iterations (including 100 EDGA micro-iterations) were implemented using the laptop, then the workstation was employed until the $30^{th}$ PE-EDGA iteration (i.e., the 200$^{th}$ EDGA micro-iteration), and finally, the HPC clusters went to the front.
Illustrations of the CI completeness and the number of ONVs during the PE-EDGA iterations are shown in Fig. \ref{fig_eight}. 
It was found that the population growth of the ONVs was quite stable (about 800-1200 were recorded in each EDGA iteration), though the computational time for each iteration was different. A final CI completeness of 0.99112 (337,268 ONVs and 94.4\% correlation energy) was obtained after 230 PE-EDGA iterations (a total of 800 EDGA micro-iterations).
It should be noted that only about 0.00000056\% ONVs were used to reach this CI completeness for a Hilbert space with a total of $\sim 6\times 10^{13}$ dimensions. 

\begin{table}[!hbp]\centering
\begin{footnotesize}
 \begin{threeparttable}
    \caption{Typical points of completeness ($\sum c_i$), number of NOVs, electronic energy (absolute energy values and percentages of correlation energy), and elapsed times during PE-EDGA iterations in constructing a CI wave function for an FDO$^{-}$ molecule. The reference HF energy and DMRG-CASSCF energy of this state are -1625.63364869 Hartree and -1626.03863532 Hartree, respectively.}\label{tab-fdo} 
  \begin{tabular}{cccccccccccc}
  \hline
  \hline
\multicolumn{2}{c}{Iteration}  && Completeness & \multirow{2}{*}{nNOVs}  & Energy & Corr. energy & Elapsed time & \multirow{2}{*}{Architecture}  \\   
(PE-EDGA) & (EDGA) && ($\sum c_i^2$) &   & (Hartree) & (\%)  & (min)  &  \\   
  \hline
   \multicolumn{3}{c}{Recorded $|c_i| > 0.0001$} \\
  0  & 0 && 0.75002 & 921 &  -1625.72485319  & 22.5\% &- & Laptop$^a$ \\ 
  1  & 5 && 0.89476 & 11,716 & -1625.89714375 & 65.1\% &$<$ 1 & Laptop$^a$\\  
  3  & 15 && 0.93079 & 19,591 & -1625.93085002 & 73.4\% & +2 & Laptop$^a$\\  
  6  & 30 && 0.94588 & 31,907 &  -1625.94803258 & 77.6\% &  +9 & Laptop$^a$ \\    
  10  & 50 && 0.95802 & 57,119 &  -1625.96046148 & 80.7\% & +10 & Laptop$^a$ \\ 
  15  & 75 && 0.96854 & 102,721 &  -1625.97427780 & 84.1\% &  +62 & Laptop$^a$ \\  
  20  & 100 && 0.97274 & 131,654 &  -1625.98025874 & 85.6\% & +50 & Laptop$^a$ \\ 
  25  & 150 && 0.97848 & 179,898 & -1625.98970877 & 87.9\% & +80 & Workstation$^b$ \\ 
  30  & 200 && 0.98184 & 211,475 & -1625.99641935 & 89.6\% & +139 & Workstation$^b$ \\ 
  80  & 350 && 0.98720 & 273,087 & -1626.00807037 & 92.4\% & +30 & HPC cluster-2$^c$ \\   
 130  & 500 && 0.98936 & 306,605 & -1626.01243486 & 93.5\% & +33 & HPC cluster-2$^c$ \\ 
 180  & 650 && 0.99049 & 325,321 & -1626.01462559 & 94.1\% & +35 & HPC cluster-2$^c$ \\   
 230  & 800 && 0.99112 & 337,268 &  -1626.01596185 & 94.4\% & +36 & HPC cluster-2$^c$ \\    
    \multicolumn{3}{c}{Recorded $|c_i| > 0.00005$} \\
 280  & 950 && 0.99351 & 717,543 &  -1626.02130479 & 95.7\% & +63 & HPC cluster-2$^c$ \\  
 330  & 1100 && 0.99416 & 784,622 &  -1626.02275378 & 96.1\% & +69 & HPC cluster-2$^c$ \\  
    \multicolumn{3}{c}{Recorded $|c_i| > 0.00002$} \\
 350  & 1140 && 0.995214 & 1,661,846 & -1626.02532460 & 96.7\%  & +39 & HPC cluster-2$^d$ \\     
  \hline
  \hline
  \end{tabular}
\scriptsize{$^a$ 5 EDGA micro-iterations in each PE-EDGA iteration using 6 PEs and 6 chares.}\\
\scriptsize{$^b$ 10 EDGA micro-iterations in each PE-EDGA iteration using 24 PEs and 24 chares.}\\
\scriptsize{$^c$ 3 EDGA micro-iterations in each PE-EDGA iteration using 960 PEs and 960 chares.}\\
\scriptsize{$^d$ 2 EDGA micro-iterations in each PE-EDGA iteration using 1920 PEs and 1920 chares.}\\
 \end{threeparttable}
\end{footnotesize}
\end{table}

\begin{figure}[htbp]
\begin{minipage}[t]{0.1\linewidth}
    \centering
	\includegraphics[height=5.625cm, width=9.0cm]{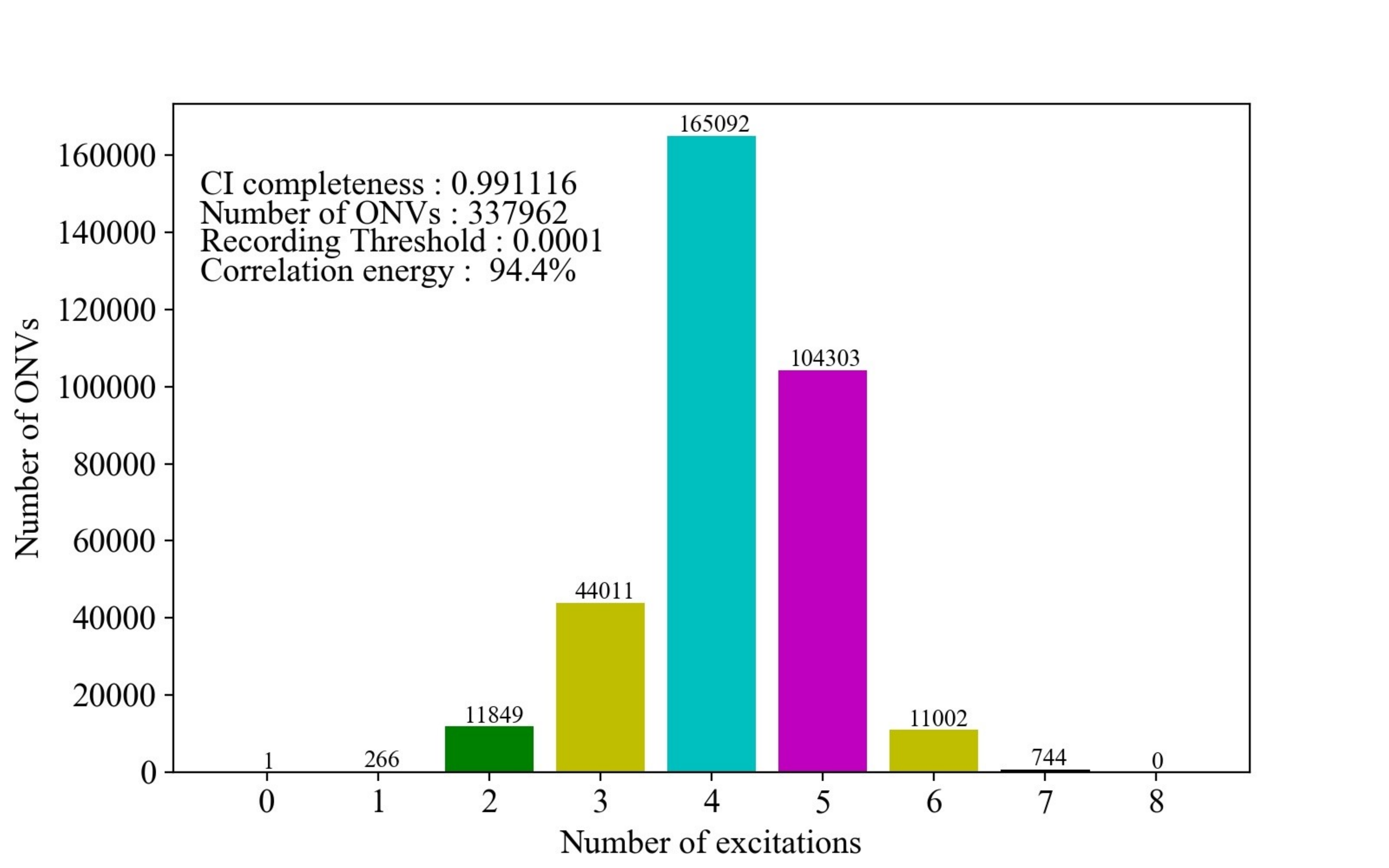}
	\end{minipage}
	\hfill
	\begin{minipage}[t]{0.5\linewidth}
	\centering
	\includegraphics[height=5.625cm,width=9.0cm]{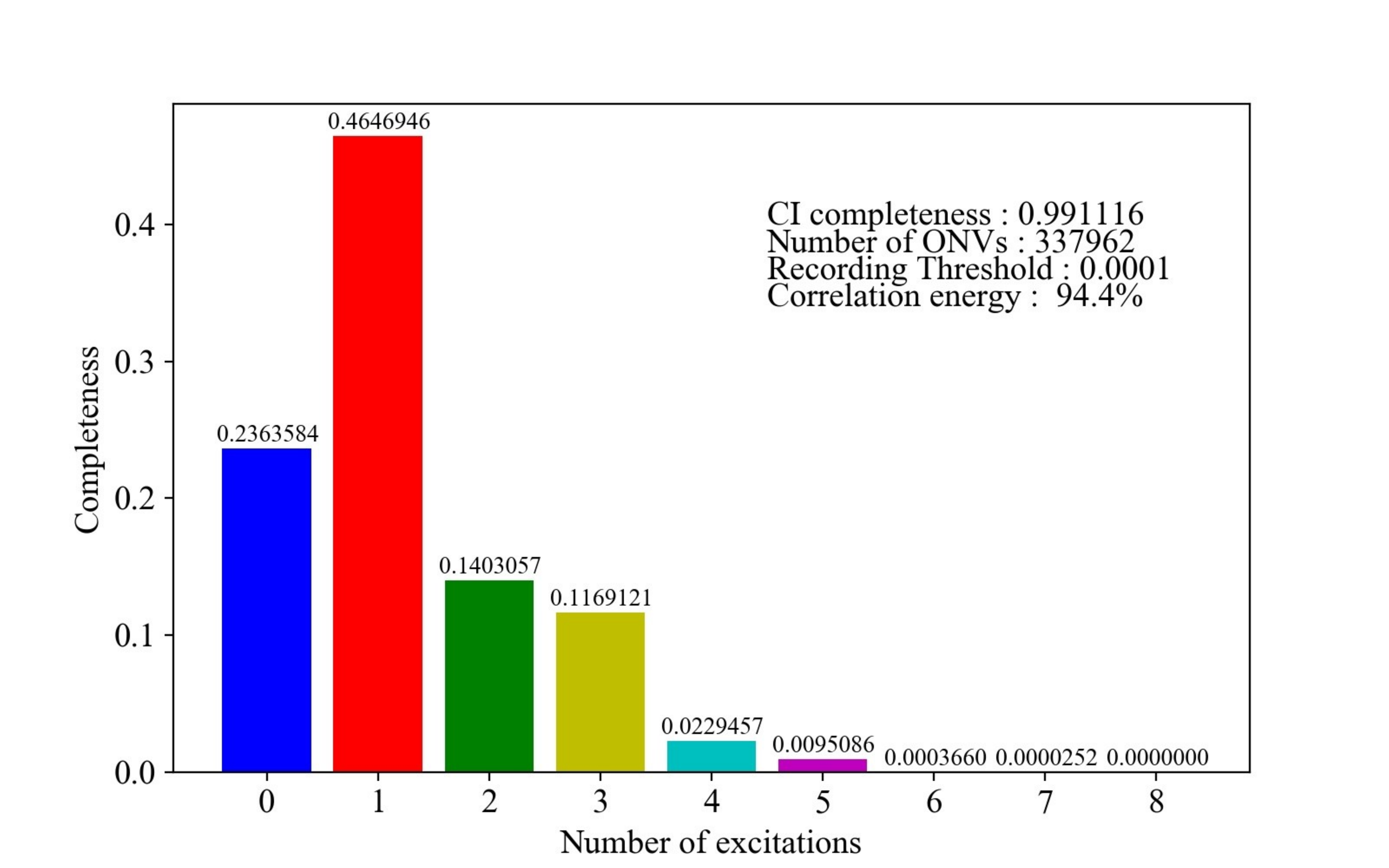}
	\end{minipage}
	
	\begin{minipage}[t]{0.1\linewidth}
    \centering
	\includegraphics[height=5.625cm, width=9.0cm]{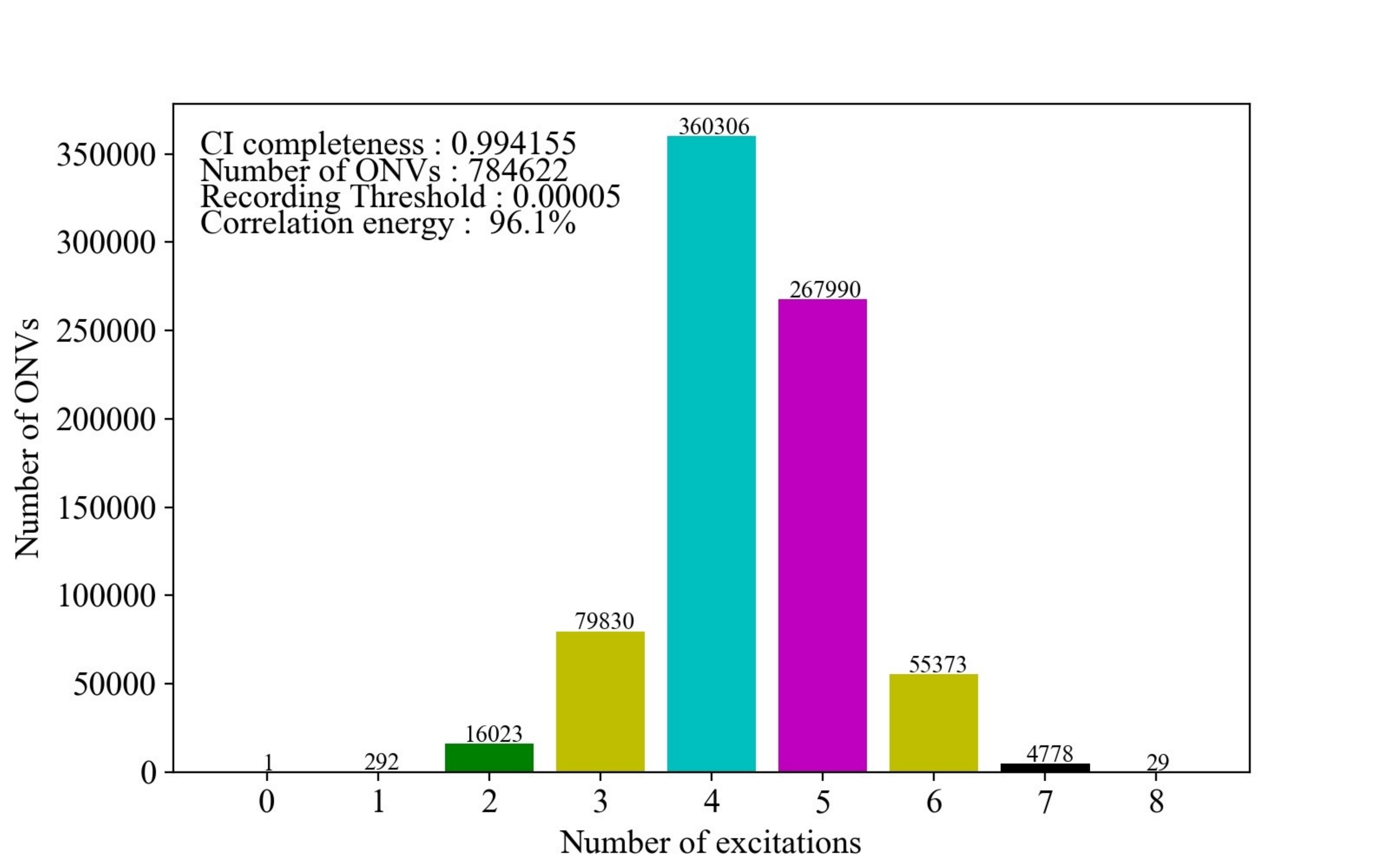}
	\end{minipage}
	\hfill
	\begin{minipage}[t]{0.5\linewidth}
	\centering
	\includegraphics[height=5.625cm,width=9.0cm]{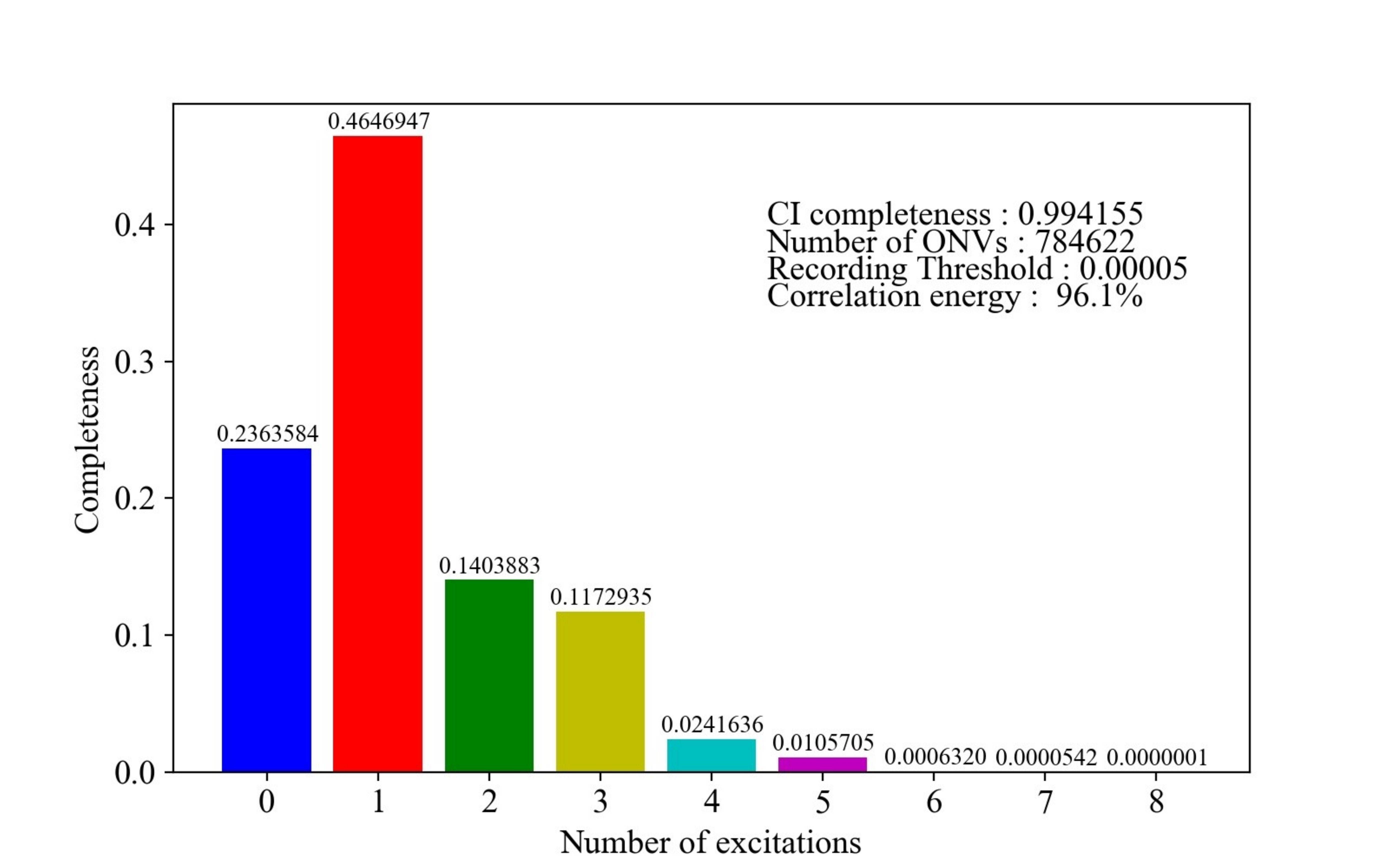}
	\end{minipage}	
	
	\begin{minipage}[t]{0.1\linewidth}
    \centering
	\includegraphics[height=5.625cm, width=9.0cm]{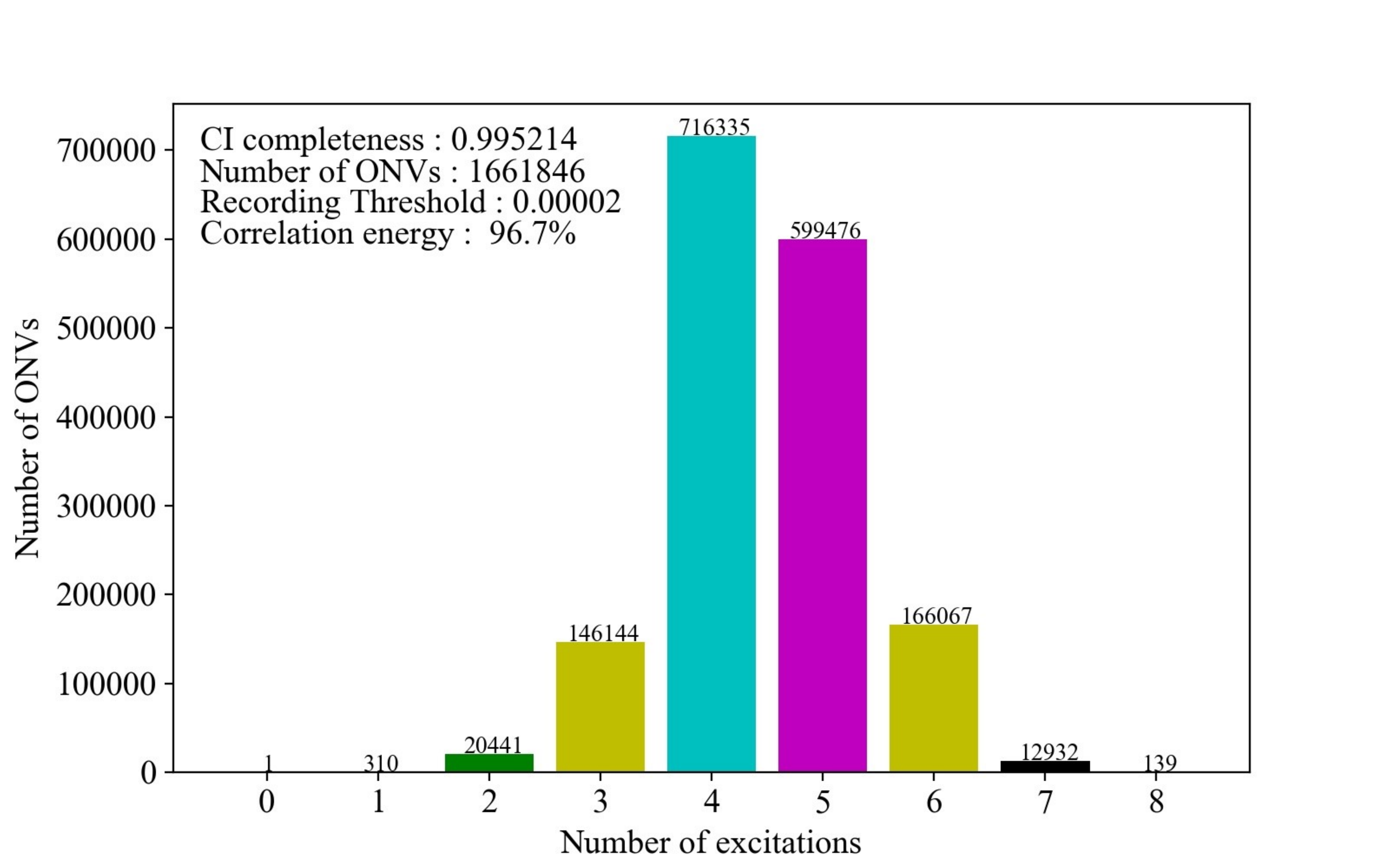}
	\end{minipage}
	\hfill
	\begin{minipage}[t]{0.5\linewidth}
	\centering
	\includegraphics[height=5.625cm,width=9.0cm]{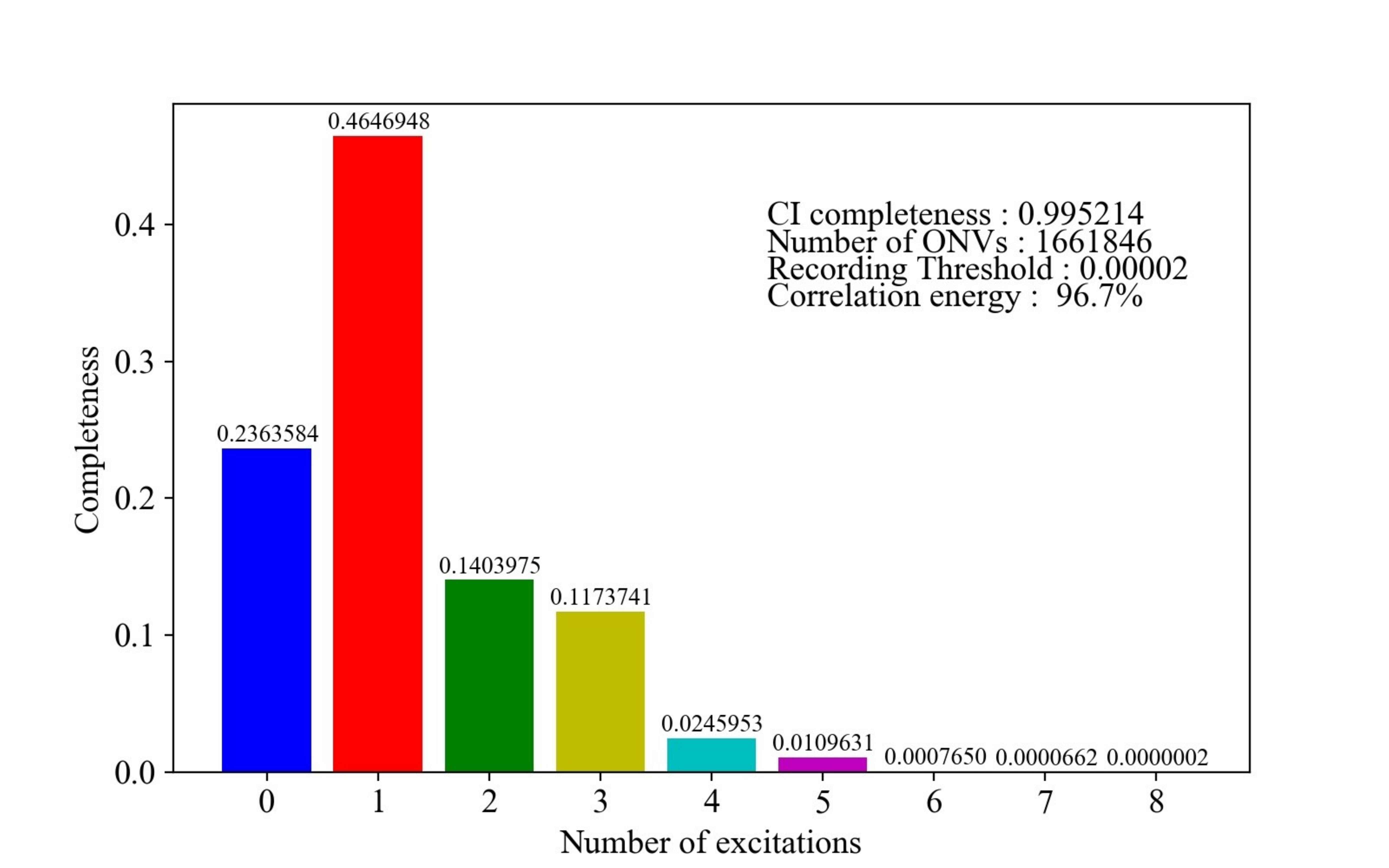}
	\end{minipage}	
	
	\caption{\label{tab-ws} Excitation analysis of the number of ONVs and CI completeness for the different number of excitations of the reconstructed CI expansion of FDO$^-$ molecule using the (30e,26o) active space. Top: after 230 PE-EDGA (i.e. 800 EDGA) iterations; middle: after 330 PE-EDGA (i.e. 1100 EDGA) iterations; bottom: after 350 PE-EDGA (i.e. 1140 EDGA) iterations}
	\label{fig_nine}
\end{figure}

Typical points of iterations, completeness, number of ONVs, electronic energies, and the elapsed times are listed in Table.\ref{tab-fdo}. Both the completeness and the number of ONVs increased rapidly (e.g., a result of $>$0.958 CI completeness [57,119 ONVs and 80.8\% correlation energy]) was obtained within 20 min using the laptop, and finally, a result of $>$0.991 CI completeness [337,268 ONVs and 94.4\% correlation energy] was obtained with the recording threshold $0.0001$). At this point, the energy derivation between the sampled state and reference DMRG state was about 22.7 mHartree. When the threshold $0.00005$ was used, the derivation turned to 15.9 mHartree and recovered 96.1\% correlation energy. The derivation was further reduced to 13.3 mHartree with a 96.7\% correlation energy when further reducing the threshold to $0.00002$.
The last updated CI expansions using different recording thresholds were also analyzed by distinguishing the excitation numbers, and the results are illustrated in Fig.\ref{fig_nine}. As can be clearly observed, the non-dynamical effects attributable to the number of ONVs and multiple excitations (i.e., $>2$ excited electrons) played an important role in the description of the electronic state. 

As previously discussed, the parallel efficiency should improve once the PE-EDGA process is employed based on the MPS-to-CI kernel. This is attributable to the automatic load balancing and the object migration facilities that could be functioned when there were several asynchronous EDGA micro-iterations in each PE-EDGA iteration. The EDGA micro-iterations were asynchronously executed in {\sc Charm++}; therefore the ONVs and MPS could be dynamical distributed among the PEs (cores). As such, the efficiency improved successively, as shown in Table.\ref{tab-hpc2}. For instance, there was a stepped efficiency improvement from one to four EDGA micro-iterations in each PE-EDGA iteration ranging from 96 to 1920 PEs (cores). The typical parallel efficiency of 960 PEs (cores) improved from about 33\% (PE-EDGA with 1-EDGA) to about 49\% (PE-EDGA with 4-EDGA). The improvements of $>3840$ PEs/cores cases were not obvious, as the system sizes were too small to be handled.
However, the asynchronous execution also meant that the occupation of memory was proportional to the number of micro-iterations during each PE-EDGA iteration, which reduced the efficiency from the five EDGA micro-iterations. 
Based on these results, the more ONVs expansions there are at the start, the less micro-iterations should be used in calculations. 

\begin{table}[!hbp]\centering
\begin{footnotesize}
 \begin{threeparttable}  
  \caption{The wall time ($s$) and parallel efficiency (\%, shown in brackets) for the implementation of a PE-EDGA iteration (based on the ONVs after the $30^{th}$ PE-EDGA iteration) with different EDGA micro-iterations in HPC clusters-2. } \label{tab-hpc2} 
   \begin{tabular}{ccccccccccccccc}
  \hline
  \hline
   &&    & \multicolumn{1}{c}{PE-EDGA} &&  \multicolumn{1}{c}{PE-EDGA} &&  \multicolumn{1}{c}{PE-EDGA} &&  \multicolumn{1}{c}{PE-EDGA} && \multicolumn{1}{c}{PE-EDGA} &&\\
   \cline{4-4}   \cline{6-6}    \cline{8-8}   \cline{10-10}   \cline{12-12} 
 nPEs   &&    &   1-EDGA   &&  2-EDGA  && 3-EDGA   &&  4-EDGA  &&  5-EDGA \\   
  \hline
   96 &  && 46 &&  62  &&  79  &&  102 && 115 \\ 
  192 &  && 29 (79\%)&&  36 (86\%) &&  46 (86\%) &&  56 (91\%) && 65 (88\%)\\  
  384 &  && 20 (58\%)&&  22 (70\%) &&  27 (73\%) &&  33 (77\%) && 40 (72\%)\\  
  768 &  && 15 (38\%)&&  16 (48\%) &&  19 (52\%) &&  23 (55\%) && 28 (51\%)\\ 
  960 &  && 14 (33\%) &&  15 (41\%) &&  17 (46\%)  &&  21 (49\%) && 26 (44\%)\\ 
 1920 &  && 14 (16\%)&&  14 (22\%) &&  16 (25\%) &&  18 (28\%) && 22 (26\%)\\ 
 3840 &  && 15 (8\%)&&  16 (10\%) &&  17 (12\%) &&  18 (14\%) && 22 (13\%)\\ 
 7680 &  && 11 (5\%)&&  20  (4\%)&&  22 (4\%) &&  23  (6\%) && 26 (6\%)\\ 
 9600 &  && 12 (4\%) &&  14 (4\%) &&  24 (3\%) &&  26 (4\%) && 29 (4\%)\\ 
  \hline
  \hline
  \end{tabular}
 \end{threeparttable}
\end{footnotesize}
\end{table}

\section{Conclusion}

We presented a procedure for constructing CI expansions from MPS using the {\sc Charm++} parallel programming framework. 
This procedure was employed for the MPS-to-CI, SR-CAS, and EDGA approaches to achieve the portable parallelism.
The chare array and the PUP operations were the key factors when refactoring these approach using {\sc Charm++}.
These were only minor changes need to be adapted when basing on {\sc C++} codes.

The parallel efficiency of the MPS-to-CI kernel from 100 to 1000 cores was around 43\%. Based on the MPS-to-CI kernel, the SR-CAS and EDGA approaches could be refactored. In addition, the PE-EDGA approach, which benefited from the parallelism with population expansion during the EDGA iterations, was also presented to functionalize the {\sc Charm++} automatic load balancing and object migration facilities. It was shown that PE-EDGA could construct a stochastic CI wave function from a super large Hilbert space, and the parallel efficiency was clearly improved through asynchronous executions. For example, the typical parallel efficiency of PE-EDGA from 96 to 960 cores was improved from around 33\% to about 49\% when adjusting the number of EDGA micro-iterations in each PE-EDGA iteration.

The results of the 1,2-dioxetanone and FDO$^-$ molecules demonstrated that the procedure presented could be flexibly employed among various multi-core architectures, ranging from laptop to HPC clusters. The optimum parameters for the calculations were evaluated, and it was demonstrated that the number of chares should be equal or just few multiples of the nPEs (ncores) to maximize efficiency.
Using the optimum parameters, the CAS-type CI wave functions for the FDO$^-$ molecule could be constructed using different CI recording thresholds. 
For example, the results of 3.37$\times$10$^5$ NOVs (CI completeness 0.9911 and 94.4\% correlation energy with CI recording threshold $0.0001$ using around 0.00000056\% total ONVs) and 7.84$\times$10$^5$ NOVs (0.9942, 96.1\%, $0.00005$, and 0.0000013\%) were obtained in turn.  
A final result of 1.66$\times$10$^6$ ONVs (0.9952, 96.7\%, $0.00002$, and 0.0000028\%) was obtained, and the derivation of the calculated energy between the sampling wave function and reference DMRG energy was only around 13.3 mHartree. 

The constructed CI wave functions can be successively improved simply by adding the iterations and/or modifying the recording threshold; thus, a controllable reference CI wave function could be used for further deterministic MR calculations,
such as the ec-MRCI \cite{luo2018externally} or ENPT2 \cite{song2019multi} corrections. We are currently working in this direction, particularly in coordination with heterogeneous architectures in the HPC environment. We hope that the magnitude of the calculated system can be fundamentally extended by combining advances in both quantum chemistry and computer science.

\section{Acknowledgement}
Ma thanks the fruitful discussions with Prof. Haibo Ma. The work was supported by the National Key Research and Development Program of China (No. 2016YFB0200800), National Natural Science Foundation of China (Grant No. 21703260), Knowledge Innovation Program of the Chinese Academy of Sciences (No. XXH13506-302,402,403), and Strategic Priority Research Programme (XDC01040000).

\bibliography{EDGA_charmpp_ref}

\begin{thebibliography}{76}%
\makeatletter
\providecommand \@ifxundefined [1]{%
 \@ifx{#1\undefined}
}%
\providecommand \@ifnum [1]{%
 \ifnum #1\expandafter \@firstoftwo
 \else \expandafter \@secondoftwo
 \fi
}%
\providecommand \@ifx [1]{%
 \ifx #1\expandafter \@firstoftwo
 \else \expandafter \@secondoftwo
 \fi
}%
\providecommand \natexlab [1]{#1}%
\providecommand \enquote  [1]{``#1''}%
\providecommand \bibnamefont  [1]{#1}%
\providecommand \bibfnamefont [1]{#1}%
\providecommand \citenamefont [1]{#1}%
\providecommand \href@noop [0]{\@secondoftwo}%
\providecommand \href [0]{\begingroup \@sanitize@url \@href}%
\providecommand \@href[1]{\@@startlink{#1}\@@href}%
\providecommand \@@href[1]{\endgroup#1\@@endlink}%
\providecommand \@sanitize@url [0]{\catcode `\\12\catcode `\$12\catcode
  `\&12\catcode `\#12\catcode `\^12\catcode `\_12\catcode `\%12\relax}%
\providecommand \@@startlink[1]{}%
\providecommand \@@endlink[0]{}%
\providecommand \url  [0]{\begingroup\@sanitize@url \@url }%
\providecommand \@url [1]{\endgroup\@href {#1}{\urlprefix }}%
\providecommand \urlprefix  [0]{URL }%
\providecommand \Eprint [0]{\href }%
\providecommand \doibase [0]{http://dx.doi.org/}%
\providecommand \selectlanguage [0]{\@gobble}%
\providecommand \bibinfo  [0]{\@secondoftwo}%
\providecommand \bibfield  [0]{\@secondoftwo}%
\providecommand \translation [1]{[#1]}%
\providecommand \BibitemOpen [0]{}%
\providecommand \bibitemStop [0]{}%
\providecommand \bibitemNoStop [0]{.\EOS\space}%
\providecommand \EOS [0]{\spacefactor3000\relax}%
\providecommand \BibitemShut  [1]{\csname bibitem#1\endcsname}%
\let\auto@bib@innerbib\@empty
\bibitem [{\citenamefont {Li}\ \emph {et~al.}(2005)\citenamefont {Li},
  \citenamefont {Shu}, \citenamefont {Chen}, \citenamefont {Wang},\ and\
  \citenamefont {Zheng}}]{LIAnalysis}%
  \BibitemOpen
  \bibfield  {author} {\bibinfo {author} {\bibfnamefont {J.}~\bibnamefont
  {Li}}, \bibinfo {author} {\bibfnamefont {J.}~\bibnamefont {Shu}}, \bibinfo
  {author} {\bibfnamefont {Y.}~\bibnamefont {Chen}}, \bibinfo {author}
  {\bibfnamefont {D.}~\bibnamefont {Wang}}, \ and\ \bibinfo {author}
  {\bibfnamefont {W.}~\bibnamefont {Zheng}},\ }\href@noop {} {\bibfield
  {journal} {\bibinfo  {journal} {Tsinghua Sci. Technol.}\ }\textbf {\bibinfo
  {volume} {10}},\ \bibinfo {pages} {304} (\bibinfo {year} {2005})}\BibitemShut
  {NoStop}%
\bibitem [{\citenamefont {Lee}, \citenamefont {Min},\ and\ \citenamefont
  {Eigenmann}(2009)}]{OpenMP}%
  \BibitemOpen
  \bibfield  {author} {\bibinfo {author} {\bibfnamefont {S.}~\bibnamefont
  {Lee}}, \bibinfo {author} {\bibfnamefont {S.-J.}\ \bibnamefont {Min}}, \ and\
  \bibinfo {author} {\bibfnamefont {R.}~\bibnamefont {Eigenmann}},\ }\href@noop
  {} {\bibfield  {journal} {\bibinfo  {journal} {ACM Sigplan Notices}\ }\textbf
  {\bibinfo {volume} {44}},\ \bibinfo {pages} {101} (\bibinfo {year}
  {2009})}\BibitemShut {NoStop}%
\bibitem [{\citenamefont {Smith}\ and\ \citenamefont
  {Bull}(2001)}]{JoseAnalysis}%
  \BibitemOpen
  \bibfield  {author} {\bibinfo {author} {\bibfnamefont {L.}~\bibnamefont
  {Smith}}\ and\ \bibinfo {author} {\bibfnamefont {M.}~\bibnamefont {Bull}},\
  }\href@noop {} {\bibfield  {journal} {\bibinfo  {journal} {Sci. Programming}\
  }\textbf {\bibinfo {volume} {9}},\ \bibinfo {pages} {83} (\bibinfo {year}
  {2001})}\BibitemShut {NoStop}%
\bibitem [{\citenamefont {Ayguad{\'e}}\ \emph {et~al.}(2008)\citenamefont
  {Ayguad{\'e}}, \citenamefont {Copty}, \citenamefont {Duran}, \citenamefont
  {Hoeflinger}, \citenamefont {Lin}, \citenamefont {Massaioli}, \citenamefont
  {Teruel}, \citenamefont {Unnikrishnan},\ and\ \citenamefont
  {Zhang}}]{jin1999openmp}%
  \BibitemOpen
  \bibfield  {author} {\bibinfo {author} {\bibfnamefont {E.}~\bibnamefont
  {Ayguad{\'e}}}, \bibinfo {author} {\bibfnamefont {N.}~\bibnamefont {Copty}},
  \bibinfo {author} {\bibfnamefont {A.}~\bibnamefont {Duran}}, \bibinfo
  {author} {\bibfnamefont {J.}~\bibnamefont {Hoeflinger}}, \bibinfo {author}
  {\bibfnamefont {Y.}~\bibnamefont {Lin}}, \bibinfo {author} {\bibfnamefont
  {F.}~\bibnamefont {Massaioli}}, \bibinfo {author} {\bibfnamefont
  {X.}~\bibnamefont {Teruel}}, \bibinfo {author} {\bibfnamefont
  {P.}~\bibnamefont {Unnikrishnan}}, \ and\ \bibinfo {author} {\bibfnamefont
  {G.}~\bibnamefont {Zhang}},\ }\href@noop {} {\bibfield  {journal} {\bibinfo
  {journal} {IEEE Transactions on Parallel and Distributed Systems}\ }\textbf
  {\bibinfo {volume} {20}},\ \bibinfo {pages} {404} (\bibinfo {year}
  {2008})}\BibitemShut {NoStop}%
\bibitem [{\citenamefont {Zheng}, \citenamefont {Shi},\ and\ \citenamefont
  {Kalé}(2004)}]{Zheng2004FTC}%
  \BibitemOpen
  \bibfield  {author} {\bibinfo {author} {\bibfnamefont {G.}~\bibnamefont
  {Zheng}}, \bibinfo {author} {\bibfnamefont {L.}~\bibnamefont {Shi}}, \ and\
  \bibinfo {author} {\bibfnamefont {L.~V.}\ \bibnamefont {Kalé}},\ }\bibfield
  {booktitle} {\emph {\bibinfo {booktitle} {2004 IEEE International Conference
  on Cluster Computing (CLUSTER 2004), September 20-23 2004, San Diego,
  California, USA}},\ }\href@noop {} {\ ,\ \bibinfo {pages} {93} (\bibinfo
  {year} {2004})}\BibitemShut {NoStop}%
\bibitem [{\citenamefont {Jacobsen}, \citenamefont {Thibault},\ and\
  \citenamefont {Senocak}(2010)}]{articleJA}%
  \BibitemOpen
  \bibfield  {author} {\bibinfo {author} {\bibfnamefont {D.}~\bibnamefont
  {Jacobsen}}, \bibinfo {author} {\bibfnamefont {J.}~\bibnamefont {Thibault}},
  \ and\ \bibinfo {author} {\bibfnamefont {I.}~\bibnamefont {Senocak}},\
  }\bibfield  {booktitle} {\emph {\bibinfo {booktitle} {48th AIAA Aerospace
  Sciences Meeting Including the New Horizons Forum and Aerospace
  Exposition}},\ }\href@noop {} {\ ,\ \bibinfo {pages} {522} (\bibinfo {year}
  {2010})}\BibitemShut {NoStop}%
\bibitem [{\citenamefont {Negara}\ \emph {et~al.}(2010)\citenamefont {Negara},
  \citenamefont {Zheng}, \citenamefont {Pan}, \citenamefont {Negara},\ and\
  \citenamefont {Ricker}}]{Negara2010Automatic}%
  \BibitemOpen
  \bibfield  {author} {\bibinfo {author} {\bibfnamefont {S.}~\bibnamefont
  {Negara}}, \bibinfo {author} {\bibfnamefont {G.}~\bibnamefont {Zheng}},
  \bibinfo {author} {\bibfnamefont {K.~C.}\ \bibnamefont {Pan}}, \bibinfo
  {author} {\bibfnamefont {N.}~\bibnamefont {Negara}}, \ and\ \bibinfo {author}
  {\bibfnamefont {P.~M.}\ \bibnamefont {Ricker}},\ }in\ \href@noop {} {\emph
  {\bibinfo {booktitle} {Euro-Par 2010 Parallel Processing Workshops -
  HeteroPar, HPCC, HiBB, CoreGrid, UCHPC, HPCF, PROPER, CCPI, VHPC, Ischia,
  Italy, August 31-September 3, 2010, Revised Selected Papers}}}\ (\bibinfo
  {year} {2010})\ pp.\ \bibinfo {pages} {531--539}\BibitemShut {NoStop}%
\bibitem [{\citenamefont {Knecht}, \citenamefont {Jensen},\ and\ \citenamefont
  {Fleig}(2008)}]{Knecht2008Large}%
  \BibitemOpen
  \bibfield  {author} {\bibinfo {author} {\bibfnamefont {S.}~\bibnamefont
  {Knecht}}, \bibinfo {author} {\bibfnamefont {H.~J.~A.}\ \bibnamefont
  {Jensen}}, \ and\ \bibinfo {author} {\bibfnamefont {T.}~\bibnamefont
  {Fleig}},\ }\href@noop {} {\bibfield  {journal} {\bibinfo  {journal} {J.
  Chem. Phys.}\ }\textbf {\bibinfo {volume} {128}},\ \bibinfo {pages} {014108}
  (\bibinfo {year} {2008})}\BibitemShut {NoStop}%
\bibitem [{\citenamefont {Vogiatzis}\ \emph {et~al.}()\citenamefont
  {Vogiatzis}, \citenamefont {Ma}, \citenamefont {Olsen}, \citenamefont
  {Gagliardi},\ and\ \citenamefont {de~Jong}}]{VogiatzisPushing}%
  \BibitemOpen
  \bibfield  {author} {\bibinfo {author} {\bibfnamefont {K.~D.}\ \bibnamefont
  {Vogiatzis}}, \bibinfo {author} {\bibfnamefont {D.}~\bibnamefont {Ma}},
  \bibinfo {author} {\bibfnamefont {J.}~\bibnamefont {Olsen}}, \bibinfo
  {author} {\bibfnamefont {L.}~\bibnamefont {Gagliardi}}, \ and\ \bibinfo
  {author} {\bibfnamefont {W.~A.}\ \bibnamefont {de~Jong}},\ }\href@noop {}
  {\bibfield  {journal} {\bibinfo  {journal} {J. Chem. Phys.}\ }\textbf
  {\bibinfo {volume} {147}},\ \bibinfo {pages} {184111}}\BibitemShut {NoStop}%
\bibitem [{\citenamefont {Nickolls}\ \emph {et~al.}(2008)\citenamefont
  {Nickolls}, \citenamefont {Buck}, \citenamefont {Garland},\ and\
  \citenamefont {Skadron}}]{Takizawa2010CheCUDA}%
  \BibitemOpen
  \bibfield  {author} {\bibinfo {author} {\bibfnamefont {J.}~\bibnamefont
  {Nickolls}}, \bibinfo {author} {\bibfnamefont {I.}~\bibnamefont {Buck}},
  \bibinfo {author} {\bibfnamefont {M.}~\bibnamefont {Garland}}, \ and\
  \bibinfo {author} {\bibfnamefont {K.}~\bibnamefont {Skadron}},\ }\href@noop
  {} {\bibfield  {journal} {\bibinfo  {journal} {Queue}\ }\textbf {\bibinfo
  {volume} {6}},\ \bibinfo {pages} {40} (\bibinfo {year} {2008})}\BibitemShut
  {NoStop}%
\bibitem [{\citenamefont {Hawick}, \citenamefont {Leist},\ and\ \citenamefont
  {Playne}(2010)}]{HawickParallel}%
  \BibitemOpen
  \bibfield  {author} {\bibinfo {author} {\bibfnamefont {K.~A.}\ \bibnamefont
  {Hawick}}, \bibinfo {author} {\bibfnamefont {A.}~\bibnamefont {Leist}}, \
  and\ \bibinfo {author} {\bibfnamefont {D.}~\bibnamefont {Playne}},\
  }\href@noop {} {\bibfield  {journal} {\bibinfo  {journal} {Parallel Comput.}\
  }\textbf {\bibinfo {volume} {36}},\ \bibinfo {pages} {655} (\bibinfo {year}
  {2010})}\BibitemShut {NoStop}%
\bibitem [{\citenamefont {Ufimtsev}\ and\ \citenamefont
  {Martinez}(2008)}]{ufimtsev2008quantum}%
  \BibitemOpen
  \bibfield  {author} {\bibinfo {author} {\bibfnamefont {I.~S.}\ \bibnamefont
  {Ufimtsev}}\ and\ \bibinfo {author} {\bibfnamefont {T.~J.}\ \bibnamefont
  {Martinez}},\ }\href@noop {} {\bibfield  {journal} {\bibinfo  {journal} {J.
  Chem. Theory Comput.}\ }\textbf {\bibinfo {volume} {4}},\ \bibinfo {pages}
  {222} (\bibinfo {year} {2008})}\BibitemShut {NoStop}%
\bibitem [{\citenamefont {Asadchev}\ and\ \citenamefont
  {Gordon}(2012)}]{asadchev2012new}%
  \BibitemOpen
  \bibfield  {author} {\bibinfo {author} {\bibfnamefont {A.}~\bibnamefont
  {Asadchev}}\ and\ \bibinfo {author} {\bibfnamefont {M.~S.}\ \bibnamefont
  {Gordon}},\ }\href@noop {} {\bibfield  {journal} {\bibinfo  {journal} {J.
  Chem. Theory Comput.}\ }\textbf {\bibinfo {volume} {8}},\ \bibinfo {pages}
  {4166} (\bibinfo {year} {2012})}\BibitemShut {NoStop}%
\bibitem [{\citenamefont {Snyder~Jr}, \citenamefont {Curchod},\ and\
  \citenamefont {Mart{\'\i}nez}(2016)}]{snyder2016gpu}%
  \BibitemOpen
  \bibfield  {author} {\bibinfo {author} {\bibfnamefont {J.~W.}\ \bibnamefont
  {Snyder~Jr}}, \bibinfo {author} {\bibfnamefont {B.~F.}\ \bibnamefont
  {Curchod}}, \ and\ \bibinfo {author} {\bibfnamefont {T.~J.}\ \bibnamefont
  {Mart{\'\i}nez}},\ }\href@noop {} {\bibfield  {journal} {\bibinfo  {journal}
  {J. Phys. Chem. Lett.}\ }\textbf {\bibinfo {volume} {7}},\ \bibinfo {pages}
  {2444} (\bibinfo {year} {2016})}\BibitemShut {NoStop}%
\bibitem [{\citenamefont {Kussmann}\ and\ \citenamefont
  {Ochsenfeld}(2017)}]{kussmann2017employing}%
  \BibitemOpen
  \bibfield  {author} {\bibinfo {author} {\bibfnamefont {J.}~\bibnamefont
  {Kussmann}}\ and\ \bibinfo {author} {\bibfnamefont {C.}~\bibnamefont
  {Ochsenfeld}},\ }\href@noop {} {\bibfield  {journal} {\bibinfo  {journal} {J.
  Chem. Theory Comput.}\ }\textbf {\bibinfo {volume} {13}},\ \bibinfo {pages}
  {2712} (\bibinfo {year} {2017})}\BibitemShut {NoStop}%
\bibitem [{\citenamefont {Bernholdt}\ \emph {et~al.}(1995)\citenamefont
  {Bernholdt}, \citenamefont {Apra}, \citenamefont {Fr{\"u}chtl}, \citenamefont
  {Guest}, \citenamefont {Harrison}, \citenamefont {Kendall}, \citenamefont
  {Kutteh}, \citenamefont {Long}, \citenamefont {Nicholas}, \citenamefont
  {Nichols}, \citenamefont {Taylor}, \citenamefont {Wong}, \citenamefont
  {Fann}, \citenamefont {Littlefield},\ and\ \citenamefont
  {Nieplocha}}]{HanwellOpen}%
  \BibitemOpen
  \bibfield  {author} {\bibinfo {author} {\bibfnamefont {D.~E.}\ \bibnamefont
  {Bernholdt}}, \bibinfo {author} {\bibfnamefont {E.}~\bibnamefont {Apra}},
  \bibinfo {author} {\bibfnamefont {H.~A.}\ \bibnamefont {Fr{\"u}chtl}},
  \bibinfo {author} {\bibfnamefont {M.~F.}\ \bibnamefont {Guest}}, \bibinfo
  {author} {\bibfnamefont {R.~J.}\ \bibnamefont {Harrison}}, \bibinfo {author}
  {\bibfnamefont {R.~A.}\ \bibnamefont {Kendall}}, \bibinfo {author}
  {\bibfnamefont {R.~A.}\ \bibnamefont {Kutteh}}, \bibinfo {author}
  {\bibfnamefont {X.}~\bibnamefont {Long}}, \bibinfo {author} {\bibfnamefont
  {J.~B.}\ \bibnamefont {Nicholas}}, \bibinfo {author} {\bibfnamefont {J.~A.}\
  \bibnamefont {Nichols}}, \bibinfo {author} {\bibfnamefont {H.}~\bibnamefont
  {Taylor}}, \bibinfo {author} {\bibfnamefont {A.~T.}\ \bibnamefont {Wong}},
  \bibinfo {author} {\bibfnamefont {G.}~\bibnamefont {Fann}}, \bibinfo {author}
  {\bibfnamefont {R.~J.}\ \bibnamefont {Littlefield}}, \ and\ \bibinfo {author}
  {\bibfnamefont {J.}~\bibnamefont {Nieplocha}},\ }\href@noop {} {\bibfield
  {journal} {\bibinfo  {journal} {Int. J. Quantum Chem.}\ }\textbf {\bibinfo
  {volume} {56}},\ \bibinfo {pages} {475} (\bibinfo {year} {1995})}\BibitemShut
  {NoStop}%
\bibitem [{\citenamefont {Kendall}\ \emph {et~al.}(2000)\citenamefont
  {Kendall}, \citenamefont {Apr{\`a}}, \citenamefont {Bernholdt}, \citenamefont
  {Bylaska}, \citenamefont {Dupuis}, \citenamefont {Fann}, \citenamefont
  {Harrison}, \citenamefont {Ju}, \citenamefont {Nichols}, \citenamefont
  {Nieplocha}, \citenamefont {Straatsma}, \citenamefont {Windus},\ and\
  \citenamefont {Wong}}]{ZBH2014NWChem}%
  \BibitemOpen
  \bibfield  {author} {\bibinfo {author} {\bibfnamefont {R.~A.}\ \bibnamefont
  {Kendall}}, \bibinfo {author} {\bibfnamefont {E.}~\bibnamefont {Apr{\`a}}},
  \bibinfo {author} {\bibfnamefont {D.~E.}\ \bibnamefont {Bernholdt}}, \bibinfo
  {author} {\bibfnamefont {E.~J.}\ \bibnamefont {Bylaska}}, \bibinfo {author}
  {\bibfnamefont {M.}~\bibnamefont {Dupuis}}, \bibinfo {author} {\bibfnamefont
  {G.~I.}\ \bibnamefont {Fann}}, \bibinfo {author} {\bibfnamefont {R.~J.}\
  \bibnamefont {Harrison}}, \bibinfo {author} {\bibfnamefont {J.}~\bibnamefont
  {Ju}}, \bibinfo {author} {\bibfnamefont {J.~A.}\ \bibnamefont {Nichols}},
  \bibinfo {author} {\bibfnamefont {J.}~\bibnamefont {Nieplocha}}, \bibinfo
  {author} {\bibfnamefont {T.}~\bibnamefont {Straatsma}}, \bibinfo {author}
  {\bibfnamefont {T.~L.}\ \bibnamefont {Windus}}, \ and\ \bibinfo {author}
  {\bibfnamefont {A.~T.}\ \bibnamefont {Wong}},\ }\href@noop {} {\bibfield
  {journal} {\bibinfo  {journal} {Comput. Phys. Commun.}\ }\textbf {\bibinfo
  {volume} {128}},\ \bibinfo {pages} {260} (\bibinfo {year}
  {2000})}\BibitemShut {NoStop}%
\bibitem [{\citenamefont {Valiev}\ \emph {et~al.}(2010)\citenamefont {Valiev},
  \citenamefont {Bylaska}, \citenamefont {Govind}, \citenamefont {Kowalski},
  \citenamefont {Straatsma}, \citenamefont {Dam}, \citenamefont {Wang},
  \citenamefont {Nieplocha}, \citenamefont {Apra}, \citenamefont {Windus},\
  and\ \citenamefont {de~Jong}}]{ValievNWChem}%
  \BibitemOpen
  \bibfield  {author} {\bibinfo {author} {\bibfnamefont {M.}~\bibnamefont
  {Valiev}}, \bibinfo {author} {\bibfnamefont {E.}~\bibnamefont {Bylaska}},
  \bibinfo {author} {\bibfnamefont {N.}~\bibnamefont {Govind}}, \bibinfo
  {author} {\bibfnamefont {K.}~\bibnamefont {Kowalski}}, \bibinfo {author}
  {\bibfnamefont {T.}~\bibnamefont {Straatsma}}, \bibinfo {author}
  {\bibfnamefont {H.~V.}\ \bibnamefont {Dam}}, \bibinfo {author} {\bibfnamefont
  {D.}~\bibnamefont {Wang}}, \bibinfo {author} {\bibfnamefont {J.}~\bibnamefont
  {Nieplocha}}, \bibinfo {author} {\bibfnamefont {E.}~\bibnamefont {Apra}},
  \bibinfo {author} {\bibfnamefont {T.}~\bibnamefont {Windus}}, \ and\ \bibinfo
  {author} {\bibfnamefont {W.}~\bibnamefont {de~Jong}},\ }\href@noop {}
  {\bibfield  {journal} {\bibinfo  {journal} {Comput. Phys. Commun.}\ }\textbf
  {\bibinfo {volume} {181}},\ \bibinfo {pages} {1477} (\bibinfo {year}
  {2010})}\BibitemShut {NoStop}%
\bibitem [{\citenamefont {Isborn}\ \emph {et~al.}(2011)\citenamefont {Isborn},
  \citenamefont {Luehr}, \citenamefont {Ufimtsev},\ and\ \citenamefont
  {Mart{\'\i}nez}}]{IsbornExcited}%
  \BibitemOpen
  \bibfield  {author} {\bibinfo {author} {\bibfnamefont {C.~M.}\ \bibnamefont
  {Isborn}}, \bibinfo {author} {\bibfnamefont {N.}~\bibnamefont {Luehr}},
  \bibinfo {author} {\bibfnamefont {I.~S.}\ \bibnamefont {Ufimtsev}}, \ and\
  \bibinfo {author} {\bibfnamefont {T.~J.}\ \bibnamefont {Mart{\'\i}nez}},\
  }\href@noop {} {\bibfield  {journal} {\bibinfo  {journal} {J. Chem. Theory
  Comput.}\ }\textbf {\bibinfo {volume} {7}},\ \bibinfo {pages} {1814}
  (\bibinfo {year} {2011})}\BibitemShut {NoStop}%
\bibitem [{\citenamefont {Ufimtsev}\ and\ \citenamefont
  {Mart{\'\i}nez}(2008)}]{UfimtsevGraphical}%
  \BibitemOpen
  \bibfield  {author} {\bibinfo {author} {\bibfnamefont {I.~S.}\ \bibnamefont
  {Ufimtsev}}\ and\ \bibinfo {author} {\bibfnamefont {T.~J.}\ \bibnamefont
  {Mart{\'\i}nez}},\ }\href@noop {} {\bibfield  {journal} {\bibinfo  {journal}
  {Comput. Sci. Eng.}\ }\textbf {\bibinfo {volume} {10}},\ \bibinfo {pages}
  {26} (\bibinfo {year} {2008})}\BibitemShut {NoStop}%
\bibitem [{\citenamefont {Ufimtsev}\ and\ \citenamefont
  {Mart{\'\i}nez}(2009)}]{UfimtsevQuantum}%
  \BibitemOpen
  \bibfield  {author} {\bibinfo {author} {\bibfnamefont {I.~S.}\ \bibnamefont
  {Ufimtsev}}\ and\ \bibinfo {author} {\bibfnamefont {T.~J.}\ \bibnamefont
  {Mart{\'\i}nez}},\ }\href@noop {} {\bibfield  {journal} {\bibinfo  {journal}
  {J. Chem. Theory Comput.}\ }\textbf {\bibinfo {volume} {5}},\ \bibinfo
  {pages} {2619} (\bibinfo {year} {2009})}\BibitemShut {NoStop}%
\bibitem [{\citenamefont {Holmes}, \citenamefont {Tubman},\ and\ \citenamefont
  {Umrigar}(2016)}]{holmes2016heat}%
  \BibitemOpen
  \bibfield  {author} {\bibinfo {author} {\bibfnamefont {A.~A.}\ \bibnamefont
  {Holmes}}, \bibinfo {author} {\bibfnamefont {N.~M.}\ \bibnamefont {Tubman}},
  \ and\ \bibinfo {author} {\bibfnamefont {C.}~\bibnamefont {Umrigar}},\
  }\href@noop {} {\bibfield  {journal} {\bibinfo  {journal} {J. Chem. Theory
  Comput.}\ }\textbf {\bibinfo {volume} {12}},\ \bibinfo {pages} {3674}
  (\bibinfo {year} {2016})}\BibitemShut {NoStop}%
\bibitem [{\citenamefont {Malmqvist}, \citenamefont {Rendell},\ and\
  \citenamefont {Roos}(1990)}]{roos1990}%
  \BibitemOpen
  \bibfield  {author} {\bibinfo {author} {\bibfnamefont {P.~{\AA}.}\
  \bibnamefont {Malmqvist}}, \bibinfo {author} {\bibfnamefont {A.}~\bibnamefont
  {Rendell}}, \ and\ \bibinfo {author} {\bibfnamefont {B.~O.}\ \bibnamefont
  {Roos}},\ }\href {\doibase 10.1021/j100377a011} {\bibfield  {journal}
  {\bibinfo  {journal} {J. Phys. Chem.}\ }\textbf {\bibinfo {volume} {94}},\
  \bibinfo {pages} {5477} (\bibinfo {year} {1990})}\BibitemShut {NoStop}%
\bibitem [{\citenamefont {Ivanic}(2003)}]{ivanic2003direct}%
  \BibitemOpen
  \bibfield  {author} {\bibinfo {author} {\bibfnamefont {J.}~\bibnamefont
  {Ivanic}},\ }\href@noop {} {\bibfield  {journal} {\bibinfo  {journal} {J.
  Chem. Phys.}\ }\textbf {\bibinfo {volume} {119}},\ \bibinfo {pages} {9364}
  (\bibinfo {year} {2003})}\BibitemShut {NoStop}%
\bibitem [{\citenamefont {Gidofalvi}\ and\ \citenamefont
  {Mazziotti}(2008)}]{gidofalvi2008active}%
  \BibitemOpen
  \bibfield  {author} {\bibinfo {author} {\bibfnamefont {G.}~\bibnamefont
  {Gidofalvi}}\ and\ \bibinfo {author} {\bibfnamefont {D.~A.}\ \bibnamefont
  {Mazziotti}},\ }\href@noop {} {\bibfield  {journal} {\bibinfo  {journal} {J.
  Chem. Phys.}\ }\textbf {\bibinfo {volume} {129}},\ \bibinfo {pages} {134108}
  (\bibinfo {year} {2008})}\BibitemShut {NoStop}%
\bibitem [{\citenamefont {Cleland}, \citenamefont {Booth},\ and\ \citenamefont
  {Alavi}(2010)}]{cleland2010communications}%
  \BibitemOpen
  \bibfield  {author} {\bibinfo {author} {\bibfnamefont {D.}~\bibnamefont
  {Cleland}}, \bibinfo {author} {\bibfnamefont {G.~H.}\ \bibnamefont {Booth}},
  \ and\ \bibinfo {author} {\bibfnamefont {A.}~\bibnamefont {Alavi}},\
  }\href@noop {} {\bibfield  {journal} {\bibinfo  {journal} {J. Chem. Phys.}\
  }\textbf {\bibinfo {volume} {132}},\ \bibinfo {pages} {041103} (\bibinfo
  {year} {2010})}\BibitemShut {NoStop}%
\bibitem [{\citenamefont {Ma}, \citenamefont {Li~Manni},\ and\ \citenamefont
  {Gagliardi}(2011)}]{gagliardi2011}%
  \BibitemOpen
  \bibfield  {author} {\bibinfo {author} {\bibfnamefont {D.}~\bibnamefont
  {Ma}}, \bibinfo {author} {\bibfnamefont {G.}~\bibnamefont {Li~Manni}}, \ and\
  \bibinfo {author} {\bibfnamefont {L.}~\bibnamefont {Gagliardi}},\ }\href
  {\doibase http://dx.doi.org/10.1063/1.3611401} {\bibfield  {journal}
  {\bibinfo  {journal} {J. Chem. Phys.}\ }\textbf {\bibinfo {volume} {135}},\
  \bibinfo {pages} {044128} (\bibinfo {year} {2011})}\BibitemShut {NoStop}%
\bibitem [{\citenamefont {Petruzielo}\ \emph {et~al.}(2012)\citenamefont
  {Petruzielo}, \citenamefont {Holmes}, \citenamefont {Changlani},
  \citenamefont {Nightingale},\ and\ \citenamefont
  {Umrigar}}]{petruzielo2012semistochastic}%
  \BibitemOpen
  \bibfield  {author} {\bibinfo {author} {\bibfnamefont {F.}~\bibnamefont
  {Petruzielo}}, \bibinfo {author} {\bibfnamefont {A.}~\bibnamefont {Holmes}},
  \bibinfo {author} {\bibfnamefont {H.~J.}\ \bibnamefont {Changlani}}, \bibinfo
  {author} {\bibfnamefont {M.}~\bibnamefont {Nightingale}}, \ and\ \bibinfo
  {author} {\bibfnamefont {C.}~\bibnamefont {Umrigar}},\ }\href@noop {}
  {\bibfield  {journal} {\bibinfo  {journal} {Phys. Rev. Lett.}\ }\textbf
  {\bibinfo {volume} {109}},\ \bibinfo {pages} {230201} (\bibinfo {year}
  {2012})}\BibitemShut {NoStop}%
\bibitem [{\citenamefont {Li~Manni}\ \emph {et~al.}(2013)\citenamefont
  {Li~Manni}, \citenamefont {Ma}, \citenamefont {Aquilante}, \citenamefont
  {Olsen},\ and\ \citenamefont {Gagliardi}}]{li2013splitgas}%
  \BibitemOpen
  \bibfield  {author} {\bibinfo {author} {\bibfnamefont {G.}~\bibnamefont
  {Li~Manni}}, \bibinfo {author} {\bibfnamefont {D.}~\bibnamefont {Ma}},
  \bibinfo {author} {\bibfnamefont {F.}~\bibnamefont {Aquilante}}, \bibinfo
  {author} {\bibfnamefont {J.}~\bibnamefont {Olsen}}, \ and\ \bibinfo {author}
  {\bibfnamefont {L.}~\bibnamefont {Gagliardi}},\ }\href@noop {} {\bibfield
  {journal} {\bibinfo  {journal} {J. Chem. Theory Comput.}\ }\textbf {\bibinfo
  {volume} {9}},\ \bibinfo {pages} {3375} (\bibinfo {year} {2013})}\BibitemShut
  {NoStop}%
\bibitem [{\citenamefont {Evangelista}(2014)}]{evangelista2014adaptive}%
  \BibitemOpen
  \bibfield  {author} {\bibinfo {author} {\bibfnamefont {F.~A.}\ \bibnamefont
  {Evangelista}},\ }\href@noop {} {\bibfield  {journal} {\bibinfo  {journal}
  {J. Chem. Phys.}\ }\textbf {\bibinfo {volume} {140}},\ \bibinfo {pages}
  {124114} (\bibinfo {year} {2014})}\BibitemShut {NoStop}%
\bibitem [{\citenamefont {Thomas}\ \emph {et~al.}(2015)\citenamefont {Thomas},
  \citenamefont {Sun}, \citenamefont {Alavi},\ and\ \citenamefont
  {Booth}}]{thomas2015stochastic}%
  \BibitemOpen
  \bibfield  {author} {\bibinfo {author} {\bibfnamefont {R.~E.}\ \bibnamefont
  {Thomas}}, \bibinfo {author} {\bibfnamefont {Q.}~\bibnamefont {Sun}},
  \bibinfo {author} {\bibfnamefont {A.}~\bibnamefont {Alavi}}, \ and\ \bibinfo
  {author} {\bibfnamefont {G.~H.}\ \bibnamefont {Booth}},\ }\href@noop {}
  {\bibfield  {journal} {\bibinfo  {journal} {J. Chem. Theory Comput.}\
  }\textbf {\bibinfo {volume} {11}},\ \bibinfo {pages} {5316} (\bibinfo {year}
  {2015})}\BibitemShut {NoStop}%
\bibitem [{\citenamefont {Liu}\ and\ \citenamefont
  {Hoffmann}(2016)}]{liu2016ici}%
  \BibitemOpen
  \bibfield  {author} {\bibinfo {author} {\bibfnamefont {W.}~\bibnamefont
  {Liu}}\ and\ \bibinfo {author} {\bibfnamefont {M.~R.}\ \bibnamefont
  {Hoffmann}},\ }\href@noop {} {\bibfield  {journal} {\bibinfo  {journal} {J.
  Chem. Theory Comput.}\ }\textbf {\bibinfo {volume} {12}},\ \bibinfo {pages}
  {1169} (\bibinfo {year} {2016})}\BibitemShut {NoStop}%
\bibitem [{\citenamefont {Coe}(2018)}]{coe2018machine}%
  \BibitemOpen
  \bibfield  {author} {\bibinfo {author} {\bibfnamefont {J.~P.}\ \bibnamefont
  {Coe}},\ }\href@noop {} {\bibfield  {journal} {\bibinfo  {journal} {Journal
  of chemical theory and computation}\ }\textbf {\bibinfo {volume} {14}},\
  \bibinfo {pages} {5739} (\bibinfo {year} {2018})}\BibitemShut {NoStop}%
\bibitem [{\citenamefont {Zimmerman}\ and\ \citenamefont
  {Rask}(2019)}]{zimmerman2019evaluation}%
  \BibitemOpen
  \bibfield  {author} {\bibinfo {author} {\bibfnamefont {P.~M.}\ \bibnamefont
  {Zimmerman}}\ and\ \bibinfo {author} {\bibfnamefont {A.~E.}\ \bibnamefont
  {Rask}},\ }\href@noop {} {\bibfield  {journal} {\bibinfo  {journal} {J. Chem.
  Phys.}\ }\textbf {\bibinfo {volume} {150}},\ \bibinfo {pages} {244117}
  (\bibinfo {year} {2019})}\BibitemShut {NoStop}%
\bibitem [{\citenamefont {Tubman}\ \emph {et~al.}(2016)\citenamefont {Tubman},
  \citenamefont {Lee}, \citenamefont {Takeshita}, \citenamefont {Head-Gordon},\
  and\ \citenamefont {Whaley}}]{tubman2016deterministic}%
  \BibitemOpen
  \bibfield  {author} {\bibinfo {author} {\bibfnamefont {N.~M.}\ \bibnamefont
  {Tubman}}, \bibinfo {author} {\bibfnamefont {J.}~\bibnamefont {Lee}},
  \bibinfo {author} {\bibfnamefont {T.~Y.}\ \bibnamefont {Takeshita}}, \bibinfo
  {author} {\bibfnamefont {M.}~\bibnamefont {Head-Gordon}}, \ and\ \bibinfo
  {author} {\bibfnamefont {K.~B.}\ \bibnamefont {Whaley}},\ }\href@noop {}
  {\bibfield  {journal} {\bibinfo  {journal} {J. Chem. Phys.}\ }\textbf
  {\bibinfo {volume} {145}},\ \bibinfo {pages} {044112} (\bibinfo {year}
  {2016})}\BibitemShut {NoStop}%
\bibitem [{\citenamefont {Tubman}\ \emph {et~al.}(2018)\citenamefont {Tubman},
  \citenamefont {Freeman}, \citenamefont {Levine}, \citenamefont {Hait},
  \citenamefont {Head-Gordon},\ and\ \citenamefont
  {Whaley}}]{tubman2018modern}%
  \BibitemOpen
  \bibfield  {author} {\bibinfo {author} {\bibfnamefont {N.~M.}\ \bibnamefont
  {Tubman}}, \bibinfo {author} {\bibfnamefont {C.~D.}\ \bibnamefont {Freeman}},
  \bibinfo {author} {\bibfnamefont {D.~S.}\ \bibnamefont {Levine}}, \bibinfo
  {author} {\bibfnamefont {D.}~\bibnamefont {Hait}}, \bibinfo {author}
  {\bibfnamefont {M.}~\bibnamefont {Head-Gordon}}, \ and\ \bibinfo {author}
  {\bibfnamefont {K.~B.}\ \bibnamefont {Whaley}},\ }\href@noop {} {\bibfield
  {journal} {\bibinfo  {journal} {arXiv preprint arXiv:1807.00821}\ } (\bibinfo
  {year} {2018})}\BibitemShut {NoStop}%
\bibitem [{\citenamefont {Jin}\ \emph {et~al.}(1999)\citenamefont {Jin},
  \citenamefont {Chen}, \citenamefont {Cui}, \citenamefont {Li},\ and\
  \citenamefont {Jiang}}]{Jin1999Mechanism}%
  \BibitemOpen
  \bibfield  {author} {\bibinfo {author} {\bibfnamefont {R.~C.}\ \bibnamefont
  {Jin}}, \bibinfo {author} {\bibfnamefont {Y.~X.}\ \bibnamefont {Chen}},
  \bibinfo {author} {\bibfnamefont {W.}~\bibnamefont {Cui}}, \bibinfo {author}
  {\bibfnamefont {W.~Z.}\ \bibnamefont {Li}}, \ and\ \bibinfo {author}
  {\bibfnamefont {Y.}~\bibnamefont {Jiang}},\ }\href@noop {} {\bibfield
  {journal} {\bibinfo  {journal} {Acta Phys.-Chim. Sin.}\ }\textbf {\bibinfo
  {volume} {15}},\ \bibinfo {pages} {317} (\bibinfo {year} {1999})}\BibitemShut
  {NoStop}%
\bibitem [{\citenamefont {Schollw\"{o}ck}(2011)}]{schollwock2011}%
  \BibitemOpen
  \bibfield  {author} {\bibinfo {author} {\bibfnamefont {U.}~\bibnamefont
  {Schollw\"{o}ck}},\ }\href@noop {} {\bibfield  {journal} {\bibinfo  {journal}
  {Ann. Phys.}\ }\textbf {\bibinfo {volume} {326}},\ \bibinfo {pages} {96 }
  (\bibinfo {year} {2011})}\BibitemShut {NoStop}%
\bibitem [{\citenamefont {Chan}\ and\ \citenamefont
  {Sharma}(2011)}]{chan2011density}%
  \BibitemOpen
  \bibfield  {author} {\bibinfo {author} {\bibfnamefont {G.~K.-L.}\
  \bibnamefont {Chan}}\ and\ \bibinfo {author} {\bibfnamefont {S.}~\bibnamefont
  {Sharma}},\ }\href@noop {} {\bibfield  {journal} {\bibinfo  {journal} {Ann.
  Rev. Phys. Chem.}\ }\textbf {\bibinfo {volume} {62}},\ \bibinfo {pages} {465}
  (\bibinfo {year} {2011})}\BibitemShut {NoStop}%
\bibitem [{\citenamefont {Liu}\ \emph {et~al.}(2013{\natexlab{a}})\citenamefont
  {Liu}, \citenamefont {Kurashige}, \citenamefont {Yanai},\ and\ \citenamefont
  {Morokuma}}]{morokuma2013}%
  \BibitemOpen
  \bibfield  {author} {\bibinfo {author} {\bibfnamefont {F.}~\bibnamefont
  {Liu}}, \bibinfo {author} {\bibfnamefont {Y.}~\bibnamefont {Kurashige}},
  \bibinfo {author} {\bibfnamefont {T.}~\bibnamefont {Yanai}}, \ and\ \bibinfo
  {author} {\bibfnamefont {K.}~\bibnamefont {Morokuma}},\ }\href {\doibase
  10.1021/ct400707k} {\bibfield  {journal} {\bibinfo  {journal} {J. Chem.
  Theory Comput.}\ }\textbf {\bibinfo {volume} {9}},\ \bibinfo {pages} {4462 }
  (\bibinfo {year} {2013}{\natexlab{a}})}\BibitemShut {NoStop}%
\bibitem [{\citenamefont {Wouters}\ \emph {et~al.}(2012)\citenamefont
  {Wouters}, \citenamefont {Limacher}, \citenamefont {Van~Neck},\ and\
  \citenamefont {Ayers}}]{ayers2012}%
  \BibitemOpen
  \bibfield  {author} {\bibinfo {author} {\bibfnamefont {S.}~\bibnamefont
  {Wouters}}, \bibinfo {author} {\bibfnamefont {P.~A.}\ \bibnamefont
  {Limacher}}, \bibinfo {author} {\bibfnamefont {D.}~\bibnamefont {Van~Neck}},
  \ and\ \bibinfo {author} {\bibfnamefont {P.~W.}\ \bibnamefont {Ayers}},\
  }\href {\doibase http://dx.doi.org/10.1063/1.3700087} {\bibfield  {journal}
  {\bibinfo  {journal} {J. Chem. Phys.}\ }\textbf {\bibinfo {volume} {136}},\
  \bibinfo {pages} {134110} (\bibinfo {year} {2012})}\BibitemShut {NoStop}%
\bibitem [{\citenamefont {Sharma}\ and\ \citenamefont {Chan}(2012)}]{chan2012}%
  \BibitemOpen
  \bibfield  {author} {\bibinfo {author} {\bibfnamefont {S.}~\bibnamefont
  {Sharma}}\ and\ \bibinfo {author} {\bibfnamefont {G.~K.-L.}\ \bibnamefont
  {Chan}},\ }\href {\doibase http://dx.doi.org/10.1063/1.3695642} {\bibfield
  {journal} {\bibinfo  {journal} {J. Chem. Phys.}\ }\textbf {\bibinfo {volume}
  {136}},\ \bibinfo {pages} {124121} (\bibinfo {year} {2012})}\BibitemShut
  {NoStop}%
\bibitem [{\citenamefont {Yanai}\ \emph {et~al.}(2015)\citenamefont {Yanai},
  \citenamefont {Kurashige}, \citenamefont {Mizukami}, \citenamefont
  {Chalupsky}, \citenamefont {Lan},\ and\ \citenamefont {Saitow}}]{yana15}%
  \BibitemOpen
  \bibfield  {author} {\bibinfo {author} {\bibfnamefont {T.}~\bibnamefont
  {Yanai}}, \bibinfo {author} {\bibfnamefont {Y.}~\bibnamefont {Kurashige}},
  \bibinfo {author} {\bibfnamefont {W.}~\bibnamefont {Mizukami}}, \bibinfo
  {author} {\bibfnamefont {J.}~\bibnamefont {Chalupsky}}, \bibinfo {author}
  {\bibfnamefont {T.~N.}\ \bibnamefont {Lan}}, \ and\ \bibinfo {author}
  {\bibfnamefont {M.}~\bibnamefont {Saitow}},\ }\href@noop {} {\bibfield
  {journal} {\bibinfo  {journal} {Int. J. Quantum Chem.}\ }\textbf {\bibinfo
  {volume} {115}},\ \bibinfo {pages} {283} (\bibinfo {year}
  {2015})}\BibitemShut {NoStop}%
\bibitem [{\citenamefont {Olivares-Amaya}\ \emph {et~al.}(2015)\citenamefont
  {Olivares-Amaya}, \citenamefont {Hu}, \citenamefont {Nakatani}, \citenamefont
  {Sharma}, \citenamefont {Yang},\ and\ \citenamefont {Chan}}]{oliv15a}%
  \BibitemOpen
  \bibfield  {author} {\bibinfo {author} {\bibfnamefont {R.}~\bibnamefont
  {Olivares-Amaya}}, \bibinfo {author} {\bibfnamefont {W.}~\bibnamefont {Hu}},
  \bibinfo {author} {\bibfnamefont {N.}~\bibnamefont {Nakatani}}, \bibinfo
  {author} {\bibfnamefont {S.}~\bibnamefont {Sharma}}, \bibinfo {author}
  {\bibfnamefont {J.}~\bibnamefont {Yang}}, \ and\ \bibinfo {author}
  {\bibfnamefont {G.~K.-L.}\ \bibnamefont {Chan}},\ }\href@noop {} {\bibfield
  {journal} {\bibinfo  {journal} {J. Chem. Phys.}\ }\textbf {\bibinfo {volume}
  {142}},\ \bibinfo {pages} {034102} (\bibinfo {year} {2015})}\BibitemShut
  {NoStop}%
\bibitem [{\citenamefont {Knecht}\ \emph {et~al.}(2016)\citenamefont {Knecht},
  \citenamefont {Hedeg\aa{}rd}, \citenamefont {Keller}, \citenamefont
  {Kovyrshin}, \citenamefont {Ma}, \citenamefont {Muolo}, \citenamefont
  {Stein},\ and\ \citenamefont {Reiher}}]{knec16a}%
  \BibitemOpen
  \bibfield  {author} {\bibinfo {author} {\bibfnamefont {S.}~\bibnamefont
  {Knecht}}, \bibinfo {author} {\bibfnamefont {E.~D.}\ \bibnamefont
  {Hedeg\aa{}rd}}, \bibinfo {author} {\bibfnamefont {S.}~\bibnamefont
  {Keller}}, \bibinfo {author} {\bibfnamefont {A.}~\bibnamefont {Kovyrshin}},
  \bibinfo {author} {\bibfnamefont {Y.}~\bibnamefont {Ma}}, \bibinfo {author}
  {\bibfnamefont {A.}~\bibnamefont {Muolo}}, \bibinfo {author} {\bibfnamefont
  {C.~J.}\ \bibnamefont {Stein}}, \ and\ \bibinfo {author} {\bibfnamefont
  {M.}~\bibnamefont {Reiher}},\ }\href@noop {} {\bibfield  {journal} {\bibinfo
  {journal} {Chimia}\ }\textbf {\bibinfo {volume} {70}},\ \bibinfo {pages}
  {244} (\bibinfo {year} {2016})}\BibitemShut {NoStop}%
\bibitem [{\citenamefont {Brabec}\ \emph {et~al.}(2020)\citenamefont {Brabec},
  \citenamefont {Brandejs}, \citenamefont {Kowalski}, \citenamefont {Xantheas},
  \citenamefont {Legeza},\ and\ \citenamefont {Veis}}]{brabec2020massively}%
  \BibitemOpen
  \bibfield  {author} {\bibinfo {author} {\bibfnamefont {J.}~\bibnamefont
  {Brabec}}, \bibinfo {author} {\bibfnamefont {J.}~\bibnamefont {Brandejs}},
  \bibinfo {author} {\bibfnamefont {K.}~\bibnamefont {Kowalski}}, \bibinfo
  {author} {\bibfnamefont {S.}~\bibnamefont {Xantheas}}, \bibinfo {author}
  {\bibfnamefont {{\"O}.}~\bibnamefont {Legeza}}, \ and\ \bibinfo {author}
  {\bibfnamefont {L.}~\bibnamefont {Veis}},\ }\href@noop {} {\bibfield
  {journal} {\bibinfo  {journal} {arXiv preprint arXiv:2001.04890}\ } (\bibinfo
  {year} {2020})}\BibitemShut {NoStop}%
\bibitem [{\citenamefont {Liu}\ \emph {et~al.}(2013{\natexlab{b}})\citenamefont
  {Liu}, \citenamefont {Kurashige}, \citenamefont {Yanai},\ and\ \citenamefont
  {Morokuma}}]{LiuMultireference}%
  \BibitemOpen
  \bibfield  {author} {\bibinfo {author} {\bibfnamefont {F.}~\bibnamefont
  {Liu}}, \bibinfo {author} {\bibfnamefont {Y.}~\bibnamefont {Kurashige}},
  \bibinfo {author} {\bibfnamefont {T.}~\bibnamefont {Yanai}}, \ and\ \bibinfo
  {author} {\bibfnamefont {K.}~\bibnamefont {Morokuma}},\ }\href@noop {}
  {\bibfield  {journal} {\bibinfo  {journal} {J. Chem. Theory Comput.}\
  }\textbf {\bibinfo {volume} {9}},\ \bibinfo {pages} {4462} (\bibinfo {year}
  {2013}{\natexlab{b}})}\BibitemShut {NoStop}%
\bibitem [{\citenamefont {Angeli}, \citenamefont {Cimiraglia},\ and\
  \citenamefont {Malrieu}(2016)}]{Angelin}%
  \BibitemOpen
  \bibfield  {author} {\bibinfo {author} {\bibfnamefont {C.}~\bibnamefont
  {Angeli}}, \bibinfo {author} {\bibfnamefont {R.}~\bibnamefont {Cimiraglia}},
  \ and\ \bibinfo {author} {\bibfnamefont {J.-P.}\ \bibnamefont {Malrieu}},\
  }\href@noop {} {\bibfield  {journal} {\bibinfo  {journal} {J. Chem. Phys.}\
  }\textbf {\bibinfo {volume} {117}},\ \bibinfo {pages} {9138} (\bibinfo {year}
  {2016})}\BibitemShut {NoStop}%
\bibitem [{\citenamefont {Luo}\ \emph {et~al.}(2018)\citenamefont {Luo},
  \citenamefont {Ma}, \citenamefont {Wang},\ and\ \citenamefont
  {Ma}}]{luo2018externally}%
  \BibitemOpen
  \bibfield  {author} {\bibinfo {author} {\bibfnamefont {Z.}~\bibnamefont
  {Luo}}, \bibinfo {author} {\bibfnamefont {Y.}~\bibnamefont {Ma}}, \bibinfo
  {author} {\bibfnamefont {X.}~\bibnamefont {Wang}}, \ and\ \bibinfo {author}
  {\bibfnamefont {H.}~\bibnamefont {Ma}},\ }\href@noop {} {\bibfield  {journal}
  {\bibinfo  {journal} {J. Chem. Theory Comput.}\ }\textbf {\bibinfo {volume}
  {14}},\ \bibinfo {pages} {4747} (\bibinfo {year} {2018})}\BibitemShut
  {NoStop}%
\bibitem [{\citenamefont {Sharma}\ \emph {et~al.}(2017)\citenamefont {Sharma},
  \citenamefont {Holmes}, \citenamefont {Jeanmairet}, \citenamefont {Alavi},\
  and\ \citenamefont {Umrigar}}]{sharma2017semistochastic}%
  \BibitemOpen
  \bibfield  {author} {\bibinfo {author} {\bibfnamefont {S.}~\bibnamefont
  {Sharma}}, \bibinfo {author} {\bibfnamefont {A.~A.}\ \bibnamefont {Holmes}},
  \bibinfo {author} {\bibfnamefont {G.}~\bibnamefont {Jeanmairet}}, \bibinfo
  {author} {\bibfnamefont {A.}~\bibnamefont {Alavi}}, \ and\ \bibinfo {author}
  {\bibfnamefont {C.~J.}\ \bibnamefont {Umrigar}},\ }\href@noop {} {\bibfield
  {journal} {\bibinfo  {journal} {J. Chem. Theory Comput.}\ }\textbf {\bibinfo
  {volume} {13}},\ \bibinfo {pages} {1595} (\bibinfo {year}
  {2017})}\BibitemShut {NoStop}%
\bibitem [{\citenamefont {Song}\ \emph {et~al.}(2020)\citenamefont {Song},
  \citenamefont {Cheng}, \citenamefont {Ma},\ and\ \citenamefont
  {Ma}}]{song2019multi}%
  \BibitemOpen
  \bibfield  {author} {\bibinfo {author} {\bibfnamefont {Y.}~\bibnamefont
  {Song}}, \bibinfo {author} {\bibfnamefont {Y.}~\bibnamefont {Cheng}},
  \bibinfo {author} {\bibfnamefont {Y.}~\bibnamefont {Ma}}, \ and\ \bibinfo
  {author} {\bibfnamefont {H.}~\bibnamefont {Ma}},\ }\href@noop {} {\bibfield
  {journal} {\bibinfo  {journal} {2020, Electron. Struct.
  https://doi.org/10.1088/2516-1075/ab72db}\ } (\bibinfo {year}
  {2020})}\BibitemShut {NoStop}%
\bibitem [{\citenamefont {Moritz}\ and\ \citenamefont
  {Reiher}(2007)}]{reiher2007}%
  \BibitemOpen
  \bibfield  {author} {\bibinfo {author} {\bibfnamefont {G.}~\bibnamefont
  {Moritz}}\ and\ \bibinfo {author} {\bibfnamefont {M.}~\bibnamefont
  {Reiher}},\ }\href {\doibase http://dx.doi.org/10.1063/1.2741527} {\bibfield
  {journal} {\bibinfo  {journal} {J. Chem. Phys.}\ }\textbf {\bibinfo {volume}
  {126}},\ \bibinfo {pages} {244109} (\bibinfo {year} {2007})}\BibitemShut
  {NoStop}%
\bibitem [{\citenamefont {Boguslawski}, \citenamefont {Marti},\ and\
  \citenamefont {Reiher}(2011)}]{boguslawski2011construction}%
  \BibitemOpen
  \bibfield  {author} {\bibinfo {author} {\bibfnamefont {K.}~\bibnamefont
  {Boguslawski}}, \bibinfo {author} {\bibfnamefont {K.~H.}\ \bibnamefont
  {Marti}}, \ and\ \bibinfo {author} {\bibfnamefont {M.}~\bibnamefont
  {Reiher}},\ }\href@noop {} {\bibfield  {journal} {\bibinfo  {journal} {J.
  Chem. Phys.}\ }\textbf {\bibinfo {volume} {134}},\ \bibinfo {pages} {224101}
  (\bibinfo {year} {2011})}\BibitemShut {NoStop}%
\bibitem [{\citenamefont {Luo}\ \emph {et~al.}(2017)\citenamefont {Luo},
  \citenamefont {Ma}, \citenamefont {Liu},\ and\ \citenamefont
  {Ma}}]{luo2017efficient}%
  \BibitemOpen
  \bibfield  {author} {\bibinfo {author} {\bibfnamefont {Z.}~\bibnamefont
  {Luo}}, \bibinfo {author} {\bibfnamefont {Y.}~\bibnamefont {Ma}}, \bibinfo
  {author} {\bibfnamefont {C.}~\bibnamefont {Liu}}, \ and\ \bibinfo {author}
  {\bibfnamefont {H.}~\bibnamefont {Ma}},\ }\href@noop {} {\bibfield  {journal}
  {\bibinfo  {journal} {J. Chem. Theory Comput.}\ }\textbf {\bibinfo {volume}
  {13}},\ \bibinfo {pages} {4699} (\bibinfo {year} {2017})}\BibitemShut
  {NoStop}%
\bibitem [{\citenamefont {Phillips}\ \emph {et~al.}(2005)\citenamefont
  {Phillips}, \citenamefont {Braun}, \citenamefont {Wang}, \citenamefont
  {Gumbart}, \citenamefont {Tajkhorshid}, \citenamefont {Villa}, \citenamefont
  {Chipot}, \citenamefont {Skeel}, \citenamefont {Kale},\ and\ \citenamefont
  {Schulten}}]{phillips2005scalable}%
  \BibitemOpen
  \bibfield  {author} {\bibinfo {author} {\bibfnamefont {J.~C.}\ \bibnamefont
  {Phillips}}, \bibinfo {author} {\bibfnamefont {R.}~\bibnamefont {Braun}},
  \bibinfo {author} {\bibfnamefont {W.}~\bibnamefont {Wang}}, \bibinfo {author}
  {\bibfnamefont {J.}~\bibnamefont {Gumbart}}, \bibinfo {author} {\bibfnamefont
  {E.}~\bibnamefont {Tajkhorshid}}, \bibinfo {author} {\bibfnamefont
  {E.}~\bibnamefont {Villa}}, \bibinfo {author} {\bibfnamefont
  {C.}~\bibnamefont {Chipot}}, \bibinfo {author} {\bibfnamefont {R.~D.}\
  \bibnamefont {Skeel}}, \bibinfo {author} {\bibfnamefont {L.}~\bibnamefont
  {Kale}}, \ and\ \bibinfo {author} {\bibfnamefont {K.}~\bibnamefont
  {Schulten}},\ }\href@noop {} {\bibfield  {journal} {\bibinfo  {journal} {J.
  Comput. Chem.}\ }\textbf {\bibinfo {volume} {26}},\ \bibinfo {pages} {1781}
  (\bibinfo {year} {2005})}\BibitemShut {NoStop}%
\bibitem [{\citenamefont {Nelson}\ \emph {et~al.}(1996)\citenamefont {Nelson},
  \citenamefont {Humphrey}, \citenamefont {Gursoy}, \citenamefont {Dalke},
  \citenamefont {Kal{\'e}}, \citenamefont {Skeel},\ and\ \citenamefont
  {Schulten}}]{nelson1996namd}%
  \BibitemOpen
  \bibfield  {author} {\bibinfo {author} {\bibfnamefont {M.~T.}\ \bibnamefont
  {Nelson}}, \bibinfo {author} {\bibfnamefont {W.}~\bibnamefont {Humphrey}},
  \bibinfo {author} {\bibfnamefont {A.}~\bibnamefont {Gursoy}}, \bibinfo
  {author} {\bibfnamefont {A.}~\bibnamefont {Dalke}}, \bibinfo {author}
  {\bibfnamefont {L.~V.}\ \bibnamefont {Kal{\'e}}}, \bibinfo {author}
  {\bibfnamefont {R.~D.}\ \bibnamefont {Skeel}}, \ and\ \bibinfo {author}
  {\bibfnamefont {K.}~\bibnamefont {Schulten}},\ }\href@noop {} {\bibfield
  {journal} {\bibinfo  {journal} {Int. J. High Perform. C.}\ }\textbf {\bibinfo
  {volume} {10}},\ \bibinfo {pages} {251} (\bibinfo {year} {1996})}\BibitemShut
  {NoStop}%
\bibitem [{\citenamefont {Mandal}\ \emph {et~al.}(2017)\citenamefont {Mandal},
  \citenamefont {Kim}, \citenamefont {Mikida}, \citenamefont {Chndrasekar},
  \citenamefont {Bohm}, \citenamefont {Jain}, \citenamefont {Kale},
  \citenamefont {Martyna},\ and\ \citenamefont
  {Ismail-Beigi}}]{mandal2017parallel}%
  \BibitemOpen
  \bibfield  {author} {\bibinfo {author} {\bibfnamefont {S.}~\bibnamefont
  {Mandal}}, \bibinfo {author} {\bibfnamefont {M.}~\bibnamefont {Kim}},
  \bibinfo {author} {\bibfnamefont {E.}~\bibnamefont {Mikida}}, \bibinfo
  {author} {\bibfnamefont {K.}~\bibnamefont {Chndrasekar}}, \bibinfo {author}
  {\bibfnamefont {E.}~\bibnamefont {Bohm}}, \bibinfo {author} {\bibfnamefont
  {N.}~\bibnamefont {Jain}}, \bibinfo {author} {\bibfnamefont {L.~V.}\
  \bibnamefont {Kale}}, \bibinfo {author} {\bibfnamefont {G.~J.}\ \bibnamefont
  {Martyna}}, \ and\ \bibinfo {author} {\bibfnamefont {S.}~\bibnamefont
  {Ismail-Beigi}},\ }\href@noop {} {\bibfield  {journal} {\bibinfo  {journal}
  {Bull. Am. Phys. Soc.}\ }\textbf {\bibinfo {volume} {62}} (\bibinfo {year}
  {2017})}\BibitemShut {NoStop}%
\bibitem [{\citenamefont {Jain}\ \emph {et~al.}(2015)\citenamefont {Jain},
  \citenamefont {Bhatele}, \citenamefont {Yeom}, \citenamefont {Adams},
  \citenamefont {Miniati}, \citenamefont {Mei},\ and\ \citenamefont
  {Kale}}]{jain2015charm++}%
  \BibitemOpen
  \bibfield  {author} {\bibinfo {author} {\bibfnamefont {N.}~\bibnamefont
  {Jain}}, \bibinfo {author} {\bibfnamefont {A.}~\bibnamefont {Bhatele}},
  \bibinfo {author} {\bibfnamefont {J.-S.}\ \bibnamefont {Yeom}}, \bibinfo
  {author} {\bibfnamefont {M.~F.}\ \bibnamefont {Adams}}, \bibinfo {author}
  {\bibfnamefont {F.}~\bibnamefont {Miniati}}, \bibinfo {author} {\bibfnamefont
  {C.}~\bibnamefont {Mei}}, \ and\ \bibinfo {author} {\bibfnamefont {L.~V.}\
  \bibnamefont {Kale}},\ }in\ \href@noop {} {\emph {\bibinfo {booktitle} {2015
  IEEE International Parallel and Distributed Processing Symposium}}}\
  (\bibinfo {organization} {IEEE},\ \bibinfo {year} {2015})\ pp.\ \bibinfo
  {pages} {655--664}\BibitemShut {NoStop}%
\bibitem [{\citenamefont {Jain}\ \emph {et~al.}(2016)\citenamefont {Jain},
  \citenamefont {Bohm}, \citenamefont {Mikida}, \citenamefont {Mandal},
  \citenamefont {Kim}, \citenamefont {Jindal}, \citenamefont {Li},
  \citenamefont {Ismail-Beigi}, \citenamefont {Martyna},\ and\ \citenamefont
  {Kale}}]{jain2016openatom}%
  \BibitemOpen
  \bibfield  {author} {\bibinfo {author} {\bibfnamefont {N.}~\bibnamefont
  {Jain}}, \bibinfo {author} {\bibfnamefont {E.}~\bibnamefont {Bohm}}, \bibinfo
  {author} {\bibfnamefont {E.}~\bibnamefont {Mikida}}, \bibinfo {author}
  {\bibfnamefont {S.}~\bibnamefont {Mandal}}, \bibinfo {author} {\bibfnamefont
  {M.}~\bibnamefont {Kim}}, \bibinfo {author} {\bibfnamefont {P.}~\bibnamefont
  {Jindal}}, \bibinfo {author} {\bibfnamefont {Q.}~\bibnamefont {Li}}, \bibinfo
  {author} {\bibfnamefont {S.}~\bibnamefont {Ismail-Beigi}}, \bibinfo {author}
  {\bibfnamefont {G.~J.}\ \bibnamefont {Martyna}}, \ and\ \bibinfo {author}
  {\bibfnamefont {L.~V.}\ \bibnamefont {Kale}},\ }in\ \href@noop {} {\emph
  {\bibinfo {booktitle} {International Conference on High Performance
  Computing}}}\ (\bibinfo {organization} {Springer},\ \bibinfo {year} {2016})\
  pp.\ \bibinfo {pages} {139--158}\BibitemShut {NoStop}%
\bibitem [{Cha()}]{Charm++Web}%
  \BibitemOpen
  \href@noop {} {\bibinfo  {journal} {http://charm.cs.uiuc.edu, accessed at
  2020.03.22}\ }\BibitemShut {NoStop}%
\bibitem [{\citenamefont {{Kal\'{e}}}\ and\ \citenamefont
  {Krishnan}(1993)}]{CHARM++:}%
  \BibitemOpen
\bibfield  {journal} {  }\bibfield  {author} {\bibinfo {author} {\bibfnamefont
  {L.}~\bibnamefont {{Kal\'{e}}}}\ and\ \bibinfo {author} {\bibfnamefont
  {S.}~\bibnamefont {Krishnan}},\ }\bibfield  {booktitle} {\emph {\bibinfo
  {booktitle} {{Proceedings of OOPSLA'93}}},\ }\href@noop {} {\ ,\ \bibinfo
  {pages} {91} (\bibinfo {year} {1993})}\BibitemShut {NoStop}%
\bibitem [{\citenamefont {Kale}\ and\ \citenamefont
  {Zheng}(2009)}]{Kale2009Charm}%
  \BibitemOpen
  \bibfield  {author} {\bibinfo {author} {\bibfnamefont {L.~V.}\ \bibnamefont
  {Kale}}\ and\ \bibinfo {author} {\bibfnamefont {G.}~\bibnamefont {Zheng}},\
  }\href@noop {} {\emph {\bibinfo {title} {Charm++ and AMPI: Adaptive Runtime
  Strategies via Migratable Objects}}}\ (\bibinfo  {publisher} {John Wiley \&
  Sons, Inc.},\ \bibinfo {year} {2009})\BibitemShut {NoStop}%
\bibitem [{\citenamefont {Legeza}\ and\ \citenamefont
  {S{\'o}lyom}(2004)}]{legeza2004quantum}%
  \BibitemOpen
  \bibfield  {author} {\bibinfo {author} {\bibfnamefont {{\"O}.}~\bibnamefont
  {Legeza}}\ and\ \bibinfo {author} {\bibfnamefont {J.}~\bibnamefont
  {S{\'o}lyom}},\ }\href@noop {} {\bibfield  {journal} {\bibinfo  {journal}
  {Phys. Rev. B}\ }\textbf {\bibinfo {volume} {70}},\ \bibinfo {pages} {205118}
  (\bibinfo {year} {2004})}\BibitemShut {NoStop}%
\bibitem [{\citenamefont {Rissler}, \citenamefont {Noack},\ and\ \citenamefont
  {White}(2006)}]{rissler2006measuring}%
  \BibitemOpen
  \bibfield  {author} {\bibinfo {author} {\bibfnamefont {J.}~\bibnamefont
  {Rissler}}, \bibinfo {author} {\bibfnamefont {R.~M.}\ \bibnamefont {Noack}},
  \ and\ \bibinfo {author} {\bibfnamefont {S.~R.}\ \bibnamefont {White}},\
  }\href@noop {} {\bibfield  {journal} {\bibinfo  {journal} {Chem. Phys.}\
  }\textbf {\bibinfo {volume} {323}},\ \bibinfo {pages} {519} (\bibinfo {year}
  {2006})}\BibitemShut {NoStop}%
\bibitem [{\citenamefont {Boguslawski}\ \emph {et~al.}(2012)\citenamefont
  {Boguslawski}, \citenamefont {Tecmer}, \citenamefont {Legeza},\ and\
  \citenamefont {Reiher}}]{boguslawski2012entanglement}%
  \BibitemOpen
  \bibfield  {author} {\bibinfo {author} {\bibfnamefont {K.}~\bibnamefont
  {Boguslawski}}, \bibinfo {author} {\bibfnamefont {P.}~\bibnamefont {Tecmer}},
  \bibinfo {author} {\bibfnamefont {O.}~\bibnamefont {Legeza}}, \ and\ \bibinfo
  {author} {\bibfnamefont {M.}~\bibnamefont {Reiher}},\ }\href@noop {}
  {\bibfield  {journal} {\bibinfo  {journal} {J. Phys. Chem. Lett.}\ }\textbf
  {\bibinfo {volume} {3}},\ \bibinfo {pages} {3129} (\bibinfo {year}
  {2012})}\BibitemShut {NoStop}%
\bibitem [{\citenamefont {Ma}(2020)}]{ma2020pccp}%
  \BibitemOpen
  \bibfield  {author} {\bibinfo {author} {\bibfnamefont {Y.}~\bibnamefont
  {Ma}},\ }\href {\doibase https://doi.org/10.1039/C9CP06417F} {\bibfield
  {journal} {\bibinfo  {journal} {Phys. Chem. Chem. Phys.}\ }\textbf {\bibinfo
  {volume} {22}},\ \bibinfo {pages} {4957} (\bibinfo {year}
  {2020})}\BibitemShut {NoStop}%
\bibitem [{\citenamefont {Widmark}, \citenamefont {Malmqvist},\ and\
  \citenamefont {Roos}(1990)}]{widm90}%
  \BibitemOpen
  \bibfield  {author} {\bibinfo {author} {\bibfnamefont {P.-O.}\ \bibnamefont
  {Widmark}}, \bibinfo {author} {\bibfnamefont {P.-{\AA}.}\ \bibnamefont
  {Malmqvist}}, \ and\ \bibinfo {author} {\bibfnamefont {B.~O.}\ \bibnamefont
  {Roos}},\ }\href@noop {} {\bibfield  {journal} {\bibinfo  {journal} {Theor.
  Chim. Acta}\ }\textbf {\bibinfo {volume} {77}},\ \bibinfo {pages} {291}
  (\bibinfo {year} {1990})}\BibitemShut {NoStop}%
\bibitem [{\citenamefont {Pierloot}\ \emph {et~al.}(1995)\citenamefont
  {Pierloot}, \citenamefont {Dumez}, \citenamefont {Widmark},\ and\
  \citenamefont {Roos}}]{pier95}%
  \BibitemOpen
  \bibfield  {author} {\bibinfo {author} {\bibfnamefont {K.}~\bibnamefont
  {Pierloot}}, \bibinfo {author} {\bibfnamefont {B.}~\bibnamefont {Dumez}},
  \bibinfo {author} {\bibfnamefont {P.-O.}\ \bibnamefont {Widmark}}, \ and\
  \bibinfo {author} {\bibfnamefont {B.~O.}\ \bibnamefont {Roos}},\ }\href@noop
  {} {\bibfield  {journal} {\bibinfo  {journal} {Theor. Chim. Acta}\ }\textbf
  {\bibinfo {volume} {90}},\ \bibinfo {pages} {87} (\bibinfo {year}
  {1995})}\BibitemShut {NoStop}%
\bibitem [{\citenamefont {Wolf}, \citenamefont {Reiher},\ and\ \citenamefont
  {Hess}(2002)}]{wolf02}%
  \BibitemOpen
  \bibfield  {author} {\bibinfo {author} {\bibfnamefont {A.}~\bibnamefont
  {Wolf}}, \bibinfo {author} {\bibfnamefont {M.}~\bibnamefont {Reiher}}, \ and\
  \bibinfo {author} {\bibfnamefont {B.~A.}\ \bibnamefont {Hess}},\ }\href@noop
  {} {\bibfield  {journal} {\bibinfo  {journal} {J. Chem. Phys.}\ }\textbf
  {\bibinfo {volume} {117}},\ \bibinfo {pages} {9215} (\bibinfo {year}
  {2002})}\BibitemShut {NoStop}%
\bibitem [{\citenamefont {Reiher}\ and\ \citenamefont
  {Wolf}(2004{\natexlab{a}})}]{reih04a}%
  \BibitemOpen
  \bibfield  {author} {\bibinfo {author} {\bibfnamefont {M.}~\bibnamefont
  {Reiher}}\ and\ \bibinfo {author} {\bibfnamefont {A.}~\bibnamefont {Wolf}},\
  }\href@noop {} {\bibfield  {journal} {\bibinfo  {journal} {J. Chem. Phys.}\
  }\textbf {\bibinfo {volume} {121}},\ \bibinfo {pages} {2037} (\bibinfo {year}
  {2004}{\natexlab{a}})}\BibitemShut {NoStop}%
\bibitem [{\citenamefont {Reiher}\ and\ \citenamefont
  {Wolf}(2004{\natexlab{b}})}]{reih04b}%
  \BibitemOpen
  \bibfield  {author} {\bibinfo {author} {\bibfnamefont {M.}~\bibnamefont
  {Reiher}}\ and\ \bibinfo {author} {\bibfnamefont {A.}~\bibnamefont {Wolf}},\
  }\href@noop {} {\bibfield  {journal} {\bibinfo  {journal} {J. Chem. Phys.}\
  }\textbf {\bibinfo {volume} {121}},\ \bibinfo {pages} {10945} (\bibinfo
  {year} {2004}{\natexlab{b}})}\BibitemShut {NoStop}%
\bibitem [{\citenamefont {Keller}\ \emph {et~al.}(2015)\citenamefont {Keller},
  \citenamefont {Dolfi}, \citenamefont {Troyer},\ and\ \citenamefont
  {Reiher}}]{kell15a}%
  \BibitemOpen
  \bibfield  {author} {\bibinfo {author} {\bibfnamefont {S.}~\bibnamefont
  {Keller}}, \bibinfo {author} {\bibfnamefont {M.}~\bibnamefont {Dolfi}},
  \bibinfo {author} {\bibfnamefont {M.}~\bibnamefont {Troyer}}, \ and\ \bibinfo
  {author} {\bibfnamefont {M.}~\bibnamefont {Reiher}},\ }\href@noop {}
  {\bibfield  {journal} {\bibinfo  {journal} {J. Chem. Phys.}\ }\textbf
  {\bibinfo {volume} {143}},\ \bibinfo {pages} {244118} (\bibinfo {year}
  {2015})}\BibitemShut {NoStop}%
\bibitem [{\citenamefont {Keller}\ and\ \citenamefont {Reiher}(2016)}]{kell16}%
  \BibitemOpen
  \bibfield  {author} {\bibinfo {author} {\bibfnamefont {S.}~\bibnamefont
  {Keller}}\ and\ \bibinfo {author} {\bibfnamefont {M.}~\bibnamefont
  {Reiher}},\ }\href@noop {} {\bibfield  {journal} {\bibinfo  {journal} {J.
  Chem. Phys.}\ }\textbf {\bibinfo {volume} {144}},\ \bibinfo {pages} {134101}
  (\bibinfo {year} {2016})}\BibitemShut {NoStop}%
\bibitem [{\citenamefont {Fdez.~Galv\'an}\ \emph {et~al.}(2019)\citenamefont
  {Fdez.~Galv\'an}, \citenamefont {Vacher}, \citenamefont {Alavi},
  \citenamefont {Angeli}, \citenamefont {Aquilante}, \citenamefont
  {Autschbach}, \citenamefont {Bao}, \citenamefont {Bokarev}, \citenamefont
  {Bogdanov}, \citenamefont {Carlson} \emph {et~al.}}]{fdez2019openmolcas}%
  \BibitemOpen
  \bibfield  {author} {\bibinfo {author} {\bibfnamefont {I.}~\bibnamefont
  {Fdez.~Galv\'an}}, \bibinfo {author} {\bibfnamefont {M.}~\bibnamefont
  {Vacher}}, \bibinfo {author} {\bibfnamefont {A.}~\bibnamefont {Alavi}},
  \bibinfo {author} {\bibfnamefont {C.}~\bibnamefont {Angeli}}, \bibinfo
  {author} {\bibfnamefont {F.}~\bibnamefont {Aquilante}}, \bibinfo {author}
  {\bibfnamefont {J.}~\bibnamefont {Autschbach}}, \bibinfo {author}
  {\bibfnamefont {J.~J.}\ \bibnamefont {Bao}}, \bibinfo {author} {\bibfnamefont
  {S.~I.}\ \bibnamefont {Bokarev}}, \bibinfo {author} {\bibfnamefont {N.~A.}\
  \bibnamefont {Bogdanov}}, \bibinfo {author} {\bibfnamefont {R.~K.}\
  \bibnamefont {Carlson}},  \emph {et~al.},\ }\href {\doibase
  10.1021/acs.jctc.9b00532} {\bibfield  {journal} {\bibinfo  {journal} {J.
  Chem. Theory Comput.}\ }\textbf {\bibinfo {volume} {15}},\ \bibinfo {pages}
  {5925} (\bibinfo {year} {2019})}\BibitemShut {NoStop}%
\bibitem [{git()}]{githubcode}%
  \BibitemOpen
  \href@noop {} {\bibinfo  {journal}
  {https://github.com/yingjin-ma/pe-edga-charmpp, accessed at 2020.03.22}\
  }\BibitemShut {NoStop}%
\bibitem [{bio()}]{biomed}%
  \BibitemOpen
\bibfield  {journal} {  }\href@noop {} {\bibinfo  {journal}
  {http://biomed.cngrid.org, accessed at 2020.03.22}\ }\BibitemShut {NoStop}%
\end{thebibliography}%
\end{document}